\documentclass[aps,a4paper,amsmath,twocolumn,amssymb,titlepage]{revtex4-1}
\usepackage[utf8]{inputenc}
\usepackage{amsmath,amssymb,comment}

\usepackage{graphicx}
\usepackage{braket, bbm, color, ulem}
\newcommand{\Lind}{\mathcal{L}}

\begin{document}

\title{Constructing neural stationary states for open quantum many-body systems}
\author{Nobuyuki Yoshioka} 
\email{\tt nysocloud@g.ecc.u-tokyo.ac.jp}
\author{Ryusuke Hamazaki}
\affiliation{Department of Physics, University of Tokyo, 7-3-1 Hongo, Bunkyo-ku, Tokyo 113-0033, Japan}

\begin{abstract}
We propose a new variational scheme based on the neural-network quantum states to simulate the stationary states of open quantum many-body systems.
Using the high expressive power of the variational ansatz described by the restricted Boltzmann machines, which we dub as the neural stationary state ansatz, we compute the stationary states of quantum dynamics obeying the Lindblad master equations.
The mapping of the stationary-state search problem into finding a zero-energy ground state of an appropriate Hermitian operator allows us to apply the conventional variational Monte Carlo method for the optimization.
Our method is shown to simulate various spin systems efficiently, i.e., the transverse-field Ising models in both one and two dimensions and the XYZ model in one dimension.

\end{abstract}

\date{\today}

\maketitle

\section{Introducton\label{sec:introduction}}
The dramatic development of machine learning techniques has inspired physicists to invent new numerical algorithms that further explore the frontier of  condensed matter physics~\cite{carrasquilla_2017, carleo_2017}.
Successful applications include the phase classification using the well-established algorithms such as the deep learning~\cite{leiwang_2016,carrasquilla_2017,evert_2017,hu_2017,broecker_sign_2017,schindler_2017,liu_2018,zhang_shen_2018,yoshioka_18}, 
the acceleration of Monte Carlo simulations~\cite{torlai_2016,j_liu_2017,h_li_2017,w_lei_2017,huitao_2018,yoshioka_18_2},
and the representation of the quantum many-body states using the high expressive power of the neural networks~\cite{carleo_2017,deng_2017,deng_2017_prb,gao_2017,nomura_2017,saito_2017, carleo_2018, saito_2018,glasser_2018,choo_2018, kaubruegger_2018, levine_2019, sharir_2019}.
In particular, the variational states based on the restricted Boltzmann machine (RBM) architecture have turned out to express the ground states of quantum many-body Hamiltonians composed of large number of spins efficiently, including one-dimensional (1D) and two-dimensional (2D) systems~\cite{carleo_2017} and highly entangled systems~\cite{deng_2017}.

Despite its rapid progress, however, machine learning has yet to be applied to one of the most challenging problems in modern condensed matter physics -- open quantum many-body systems.
Although the advancement of experiments~\cite{barreiro_2011,barontini_2013,labouvie_2016,tomita_2017,fitzpatrick_2017} motivates an active field of research on open quantum many-body physics, it is notoriously difficult to solve the fundamental equation of motion, which is often well captured by the Lindblad master equation~\cite{lindblad_1976}.
Due to the growth of the number of parameters in proportion to the square of the Hilbert space dimension, description of the quantum states by density matrices requires additional computational resource compared to the closed system.
Accordingly, the simulation of the Lindblad equation with the exact diagonalization method is hard even for small system sizes.
It is thus important whether the machine learning techniques help us to simulate open quantum many-body physics.
Particularly intriguing are nonequilibrium stationary states of dynamics, which can exhibit exotic structure such as
entanglement~\cite{kraus_2008,kastoryano_2011}, nontrivial topology~\cite{diehl_2011,bardyn_2013}, 
and novel dissipative phases of matter~\cite{diehl_2010,tomadin_2011,ates_2012,gong_2018,gambetta_2019}.

\begin{figure}[t]
\begin{center}
\begin{tabular}{c}
  \begin{minipage}{0.97\hsize}
    \begin{center}
     \resizebox{0.95\hsize}{!}{\includegraphics{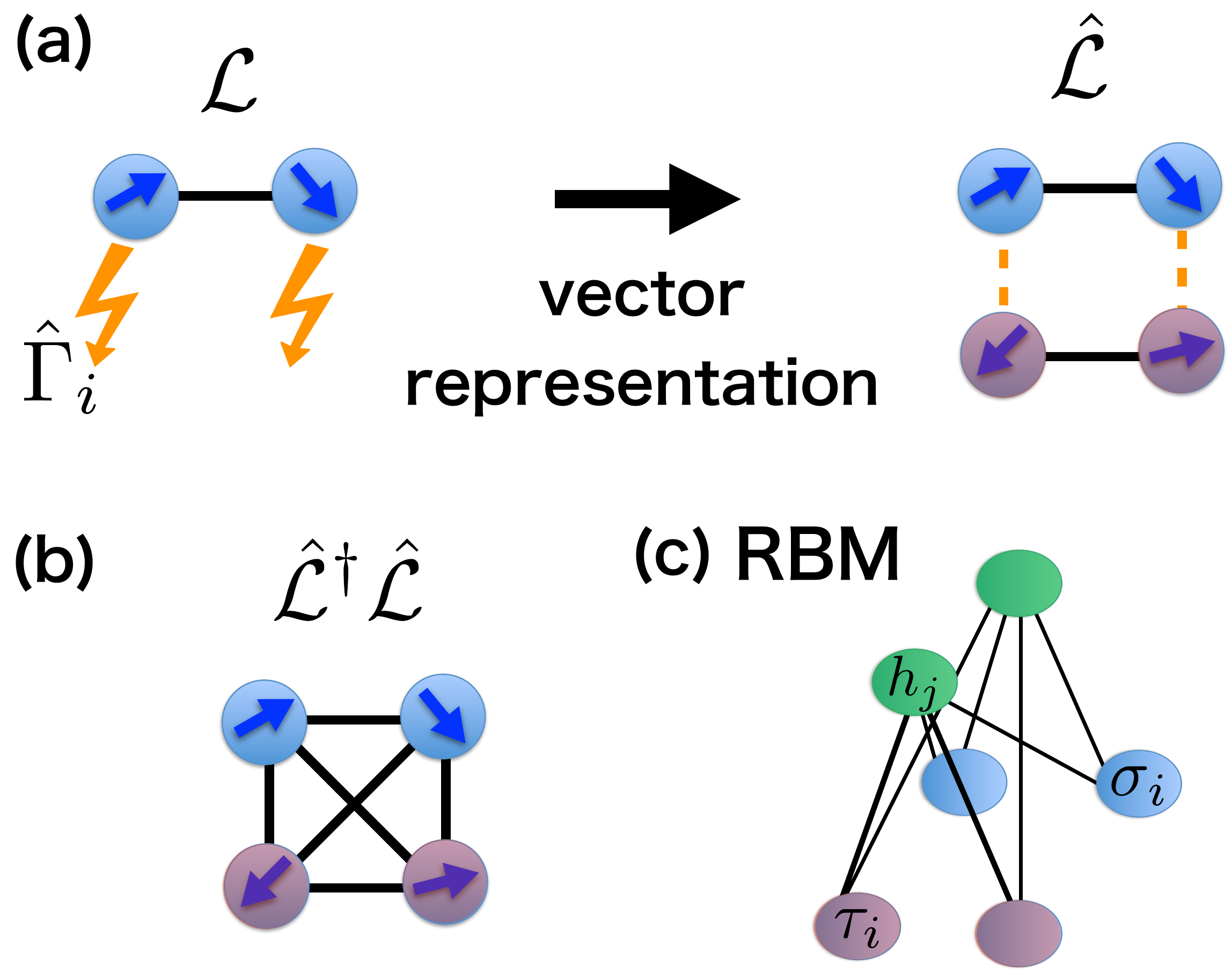}}

   \end{center}
  \end{minipage}
 
\end{tabular}
\end{center}
\caption{\label{fig:rbm} (Color Online)
Schematic illustration of our method for the case with two spins.
(a) (Left) A spin model with dissipations $\hat{\Gamma}_i$, 
\textcolor{black}{which are indicated by the yellow zig-zag arrows,} described by the Lindblad superoperator $\mathcal{L}$. The usual Hermitian interaction is denoted by the real black line.
(Right) The vector representation of the Lindblad superoperator as an operator $\mathcal{\hat{L}}$ acting on the doubled Hilbert space.
The dissipations become the non-Hermitian interactions, denoted by the yellow dotted lines,  between the physical and newly-introduced fictitious spins.
(b) The Hermitian operator $\mathcal{\hat{L}}^\dag\mathcal{\hat{L}}$ in the vector representation, whose expectation value plays a role of the cost function for the variational ansatz. 
(c) Our neural stationary state represented by the RBM. 
\textcolor{black}{The black thin lines denote the non-zero interaction parameters in the ansatz between the physical spins ${\sigma}_i$ (or fictitious spins ${\tau}_i$) and hidden spins $h_j$.}
}
\end{figure}

In this work, we present a new scheme for simulating  the stationary states of open quantum many-body systems by employing the ansatz which we refer to as the neural stationary state (NSS) in the following.
As is schematically illustrated in Fig.~\ref{fig:rbm}, our NSS method is constituted by the following three steps:
\begin{itemize}
    \item[(a)] Vector representation: Make a copy of the Hilbert space and 
    \textcolor{black}{define a non-Hermitian operator $\mathcal{\hat{L}}$, which is the generator of Lindblad dynamics in the doubled Hilbert space.}
    \item[(b)] Definition of the cost function: Consider a Hermitian positive-semidefinite operator $\mathcal{\hat{L}}^\dag \mathcal{\hat{L}}$, which becomes zero for the stationary state~\cite{cui_2015}.
    \item[(c)] Optimization: Optimize the NSS ansatz using the variational Monte Carlo method (VMC).
\end{itemize}
\textcolor{black}{We first demonstrate the expressive power of the ansatz by showing that the generic NSS exhibits volume-law entanglement entropy in the vector representation, which is the so-called operator space entanglement entropy~\cite{prosen_2007,pizorn_2009}.}
Next, we show that our NSS ansatz is capable of representing the stationary states of the dissipative transverse-field Ising models in 1D and 2D, and XYZ model in 1D.

We remark that there have been many previous proposals for simulating open quantum many-body systems numerically.
For example, the Lindblad dynamics is simulated by the density matrix renormalization group~\cite{cai_2013,werner_2016,gangat_2017,kshetrimayum_2017} under the tensor network representation, which works very well especially in 1D as long as the operator space entanglement entropy of the density matrix is small.
In addition, numerous works have focused particularly on the stationary states of the Lindblad dynamics.
Cui et al.~\cite{cui_2015} presented an elegant variational method to search for the stationary states of the Lindblad dynamics by minimizing the expectation value of $\mathcal{\hat{L}}^\dag \mathcal{\hat{L}}$ using the matrix product operator (MPO) algorithm, which is powerful for 1D systems. 
Beyond 1D, Ref.~\cite{weimer_2015} treated variational quantum states that take low-order correlations around the product states into account.
It is also notable that certain approximations beyond the mean-field theory, e.g., the cluster mean-field theory~\cite{jiasen_2016,jin_2018}, were employed.
Few methods have been proposed, however, that can efficiently capture full quantum correlations beyond 1D.

The rest of the paper is organized as follows.
A brief overview of open quantum systems in the Lindblad form and its vector representation  is given  in Sec.~\ref{sec:openquantum}. This representation allows us to map the stationary-state search problem into finding a zero-energy ground state of an appropriate Hermitian operator that is composed of the \textcolor{black}{Lindblad operator} and its Hermitian adjoint operator.
In Sec.~\ref{sec:NSS}, we introduce the NSS ansatz, which is optimized via 
\textcolor{black}{the variational Monte Carlo technique}. 
\textcolor{black}{We show in Sec.~\ref{sec:model_result} that our ansatz is capable of expressing density matrices with volume-law operator space entanglement and also} 
the stationary states of various spins systems, i.e., the transverse-field Ising models in both 1D and 2D, and the XYZ model in 1D. 
Finally, the summary of our work and the discussion on the future directions are presented in Sec.~\ref{sec:conclusions}.
\textcolor{black}{For completeness, we discuss the result for fitting random density matrices with the NSS in Appendix~\ref{sec:random_fit}, and the comparison of the computational time between the NSS and Lanczos methods is discussed in Appendix~\ref{sec:walltime}.
}

\section{Open quantum systems in the Lindblad form}\label{sec:openquantum}
In this section, we first give a brief overview of the Lindblad form for describing open quantum systems. 
To simulate the stationary states efficiently, we introduce the vector representation of mixed states.
Consequently, the stationary state of the Lindblad dynamics can be obtained by finding the ground state of an appropriate Hermitian matrix that is composed of the \textcolor{black}{Lindblad operator} and its Hermitian adjoint, or $\hat{\Lind}^{\dagger} \hat{\Lind}$. 
We propose that this problem can be solved efficiently via the conventional VMC method with insight into the optimization quality: the expectation value of 
$\hat{\Lind}^{\dagger} \hat{\Lind}$ is regarded as the cost function since the target state corresponds to the zero-energy eigenstate of it.

\subsection{Lindblad master equation}\label{subsec:lindblad}
Open quantum physics consider situations where a system interacts with their environments outside and follows non-unitary time evolution.
Such systems with certain conditions, e.g. the Markovianity, are known to be well described by the Lindblad equation~\cite{lindblad_1976}, which possesses the completely positive and trace-preserving  property. 
Concretely, the time evolution of a mixed state $\hat{\rho}(t)$ is given by
\begin{eqnarray}\label{eqn:lindblad}
\frac { d \hat{\rho} ( t ) } { d t } = {\mathcal L}\hat{\rho}(t) := - i [ \hat{H} , \hat{\rho} ( t ) ] + \sum_i\gamma_i {\mathcal D}[\hat{\Gamma}_i]\hat{\rho}(t).
\end{eqnarray}
\textcolor{black}{Here, $\mathcal{L}$ is the Lindblad superoperator, which is a linear map that takes a density matrix to another density matrix. 
The first term in the right hand side, given by the commutator $[\hat{H}, \hat{\rho}] = \hat{H}\hat{\rho} - \hat{\rho} \hat{H}$, describes the unitary dynamics ruled by the Hamiltonian $\hat{H}$. The second term describes the non-unitary dynamics due to the dissipations. The contribution of the $i$-th term, whose strength is given as $\gamma_i$, is governed by a superoperator $\mathcal{D}[\hat{\Gamma}_i]$ acting on the density matrix $\hat{\rho}(t)$ as}
\begin{eqnarray}\label{eqn:superoperator}
{\mathcal D}[\hat{\Gamma}_i]\hat{\rho}(t) &=&  \hat{\Gamma}_i \hat{\rho} ( t ) \hat{\Gamma}_i ^ { \dagger } - 
\frac { 1 } { 2 } \hat{\Gamma} _ { i } ^ { \dagger } \hat{\Gamma} _ { i } \hat{\rho} ( t ) - \frac{1}{2}\hat{\rho}(t) \hat{\Gamma}_i^{\dagger}\hat{\Gamma}_i.
\end{eqnarray}
\textcolor{black}{
 Here, $\hat{\Gamma}_i$, or the $i$-th jump operator, determines the detail of the dissipations. }

\subsection{Vector representation of the Lindblad equation}
It is known that a time-independent Lindblad equation has at least one stationary state satisfying
\begin{eqnarray}
\mathcal{L}\hat{\rho}_{\rm SS} = 0,
\end{eqnarray}
where $\hat{\rho}_{\rm SS}$ is a density matrix of the stationary state~\cite{rivas_2011}. 
To employ well-established numerical calculation schemes, we first map the density \textcolor{blue}{\sout{``}}matrix\textcolor{blue}{\sout{"}} $\hat{\rho}$ to an element in the so-called ``operator space" as $\ket{\rho}\rangle \in \mathcal{H}\otimes \mathcal{H}$.
The new representation of the state, which we call the ``vector" representation throughout this manuscript, is explicitly given by
\begin{eqnarray}\label{eqn:vector_map}
\hat{\rho} = \sum_{\sigma\tau} \rho_{\sigma\tau} \ket{\sigma}\bra{\tau}\ \ \mapsto \ket{\rho}\rangle = \frac{1}{C}\sum_{\sigma\tau}\rho_{\sigma\tau}\ket{\sigma,\tau}\rangle,
\end{eqnarray}
where $\ket{\sigma,\tau}\rangle=\ket{\sigma} \otimes \ket{\tau}\in \mathcal{H}\otimes \mathcal{H}$ is a spin configuration basis that spans $ \mathcal{H}\otimes \mathcal{H}$
and $C = \sqrt{\sum_{\sigma\tau}|\rho_{\sigma\tau}|^2}$ denotes the normalization factor. 
\textcolor{black}{In the doubled Hilbert space, we discriminate the spins denoted by $\sigma$ and $\tau$ by referring to them as the physical and fictitious spins, respectively.}
We note that the normalizations in two representations are different from each other; the trace of the matrix is set to unity, i.e., $\sum_\sigma \rho_{\sigma\sigma} = 1$, in the matrix representation,  whereas the $L^2$-norm of $\ket{\rho}\rangle$, or $\langle\braket{\rho}\rangle$, is unity in the vector representation.  

Using the mapping in Eq. (\ref{eqn:vector_map}), operators $\hat{A}$ and $\hat{B}$ that respectively act on $\hat{\rho}$ from left and right are mapped as follows,
\begin{eqnarray}\label{eqn:left_right_operator}
\hat{A}\hat{\rho}\hat{B} &=& \sum_{\sigma\mu\nu\tau} A_{\sigma\mu}\rho_{\mu\nu}B_{\nu\tau} \ket{\sigma} \bra{\tau},  \\
\mapsto |A\rho B\rangle\rangle &=& \frac{1}{C'}\sum_{\sigma\mu\nu\tau} A_{\sigma\mu} \rho_{\mu\nu} (B^T)_{\tau\nu} \ket{\sigma,\tau}\rangle\\
 &=& \hat{A} \otimes \hat{B}^T \ket{\rho}\rangle,
\end{eqnarray}
where $C'=\sum_{\sigma\tau}\left|\sum_{\mu\nu}A_{\sigma\mu}\rho_{\mu\nu}B_{\nu\tau}\right|^2$ is a normalization factor.
Applying the mapping in Eq. (\ref{eqn:left_right_operator}) to Eq. (\ref{eqn:lindblad}), we obtain the vector representation of the Lindblad equation as
\begin{align}
 \frac{d \ket{\rho(t)}\rangle}{dt} &= \hat{\mathcal{L}}\ket{\rho(t)}\rangle \nonumber\\
&=
\left(-i(\hat{H}\otimes \mathbbm{1} - \mathbbm{1}\otimes \hat{H}^T) + \sum_i \gamma_i \hat{\mathcal{D}}[\hat{\Gamma}_i]\right)\ket{\rho(t)}\rangle
\label{eqn:liouvillian_vector}
\end{align}
with
\begin{align}
\hat{\mathcal{D}}[\hat{\Gamma}_i] =  \hat{\Gamma}_i \otimes \hat{\Gamma}_i^* - \frac{1}{2}\hat{\Gamma}_i^{\dagger}\hat{\Gamma}_i \otimes \mathbbm{1} - \mathbbm{1} \otimes \frac{1}{2}\hat{\Gamma}_i^T\hat{\Gamma}_i^*.
\end{align}
Here, the 
\textcolor{black}{Lindblad operator} is denoted by the operator $\hat{\Lind}$ acting on the operator space $\mathcal{H}\otimes \mathcal{H}$. 
\textcolor{black}{Hence, in } this new representation the problem of finding the stationary state is expressed in terms of the  standard linear algebra. 
Our goal is, concretely, to solve the equation for a non-Hermitian operator $\hat{\mathcal{L}}$ as follows:
\begin{eqnarray}\label{eqn:NESS_vector}
\hat{\mathcal{L}}|\rho_{\rm SS}\rangle\rangle = 0,
\end{eqnarray}
where $\ket{\rho_{\rm SS}} \rangle$ denotes the vector representation of the stationary state.
Note that other 
\textcolor{black}{right eigenvectors} with non-zero eigenvalues satisfy
\begin{eqnarray}
\hat{\mathcal{L}}\ket{\rho_n}\rangle = \lambda_n \ket{\rho_n}\rangle,\\
e^{\hat{\mathcal{L}}t} \ket{\rho_n}\rangle = e^{\lambda_n t}\ket{\rho_n}\rangle,
\end{eqnarray}
where $\ket{\rho_n}\rangle$ is the 
with a \textcolor{black}{right} eigenvalue $\lambda_n$.
For eigenmodes that are not stationary states, 
\textcolor{black}{the real part of} the corresponding \textcolor{black}{right} eigenvalues satisfy ${\Re}[\lambda_n] < 0$ \cite{breuer_2002, rivas_2011}, which implies that the modes eventually decay. 

\subsection{Stationary state as a ``ground state" of $\hat{\mathcal{L}}^{\dagger} \hat{\mathcal{L}}$}
\label{subsec:variationalNESS}
The 
\textcolor{black}{Lindblad operator} in Eq. (\ref{eqn:NESS_vector}) is a non-Hermitian matrix, whose eigenvalues are in general complex.
In contrast, the product with the Hermitian-conjugated 
\textcolor{black}{Lindblad operator}, $\hat{\mathcal{L}}^{\dagger} \hat{\mathcal{L}}$, is a Hermitian matrix and has a real-valued non-negative spectrum. 
In this case, the lowest eigenstate(s) with eigenvalue(s) $\lambda = 0$ of $\hat{\mathcal{L}}^{\dagger} \hat{\mathcal{L}}$ correspond to the stationary state(s). In other words, $\ket{\rho_{\rm SS}}\rangle $ satisfies
\begin{eqnarray}\label{eqn:NESS_vector_withconj}
\hat{\mathcal{L}}^{\dagger} \hat{\mathcal{L}} \ket{\rho_{\rm SS}}\rangle = 0.
\end{eqnarray}
This allows us to apply the well-established ground-state search technique in closed systems such as the variational approaches, in addition to the Lanczos method,  if the  first excited energy of $\hat{\mathcal{L}}^{\dagger} \hat{\mathcal{L}}$ does not vanish~\cite{cui_2015}.
Therefore, the expectation value of $\hat{\Lind}^{\dagger} \hat{\Lind}$ is suited for the cost function in the VMC method.

\textcolor{black}{
Note that the uniqueness of the stationary states is confirmed in various systems.
For example, if the annihilation operator, or the incoherent spin flip along the $z$-axis in the language of spins, is included as the dissipation for each site, the quantum system has a unique stationary state regardless of the Hamiltonian~\cite{schirmer_2010}. 
Unique stationary states also appear for other types of dissipations, as demonstrated in, e.g., Refs.~\cite{prosen_2012,cai_2013,horstmann_2013}.
}

Let us emphasize  that the variational approach has advantage in the sense of the cost function $\langle \braket{\hat{\Lind}^{\dagger} \hat{\mathcal{L}}} \rangle$, where $\langle\braket{\hat{O}}\rangle$ denotes the expectation value of the operator $\hat{O}$ in the vector representation. 
Since $\langle \braket{\hat{\Lind}^{\dagger} \hat{\mathcal{L}}} \rangle$ is exactly zero by construction for stationary states,
this indicates the quality of the optimization.
Note the difference from usual variational problems of finding ground states of Hamiltonians, for which quantification of the optimization quality is difficult without knowing the ground-state energy. In that case, the convergence of the cost function may both yield the desired state or indicate an ill result due to the local minima.

\section{Neural Stationary States}\label{sec:NSS}
In this section, we present the method to compute our NSS for a given Lindbladian of the system. First, we describe the complex-valued RBM $\ket{\rho_{\rm RBM}}\rangle$, which introduces auxiliary binary degrees of freedom to extend expressive power of the ansatz, as follows:
\begin{align}\label{eqn:def_RBM}
\langle\braket{\sigma,\tau|\rho_{\rm RBM}}\rangle = \frac{1}{Z}&\sum_{\{h_j\}} \exp\left(\sum_{ij}W_{ij}\sigma_i h_j + \bar{W}_{ij}\tau_i h_j  \right)\nonumber\\
&\times \exp\left( \sum_i a_i \sigma_i + \bar{a}_i \tau_i + \sum_j b_j h_j \right),
\end{align}
where $W_{ij}$ ($\bar{W}_{ij}$) denotes complex interaction amplitude between the $i$-th physical (fictitious) spin $\sigma_i$ ($\tau_i$) and  $j$-th hidden spin $h_j$, $a_i$ ($\bar{a}_i$) is a complex magnetic field on the $i$-th physical (fictitious) spin, and $b_j$ is a complex magnetic field on the $j$-th hidden spin. The normalization factor $Z$ is determined such that $\langle \braket{\rho_{\rm RBM}| \rho_{\rm RBM}}\rangle = 1$. Denoting the number of the physical, fictitious, and hidden spins as $N, \bar{N}( = N),$ and $M$, respectively, we define the number ratio of the spins as $\alpha = M/(N+\bar{N})$ \textcolor{black}{to compare the performance of the NSS ansatz under different system sizes}.

Although the NSS obtained from Eq.~(\ref{eqn:def_RBM}) is not positive-semidefinite or Hermitian in general, sufficient optimization of the cost function is expected to ensure these two conditions in an approximated way~\cite{cui_2015}. In fact, we have  confirmed that absolute values of unphysical negative eigenvalues, if any, 
and 
\textcolor{black}{$||\hat{\rho}_{\rm RBM} - \hat{\rho}_{\rm RBM}^{\dagger}||/||\hat{\rho}_{\rm RBM} + \hat{\rho}_{\rm RBM}^{\dagger}||$
are in the order of $10^{-3}$.
Both quantities are sufficiently small compared to unity, which indicates that the NSS method works well, and can be further reduced by, for instance, taking larger $\alpha$.}
In the following,  physical observables such as the entropy are computed using the symmetrized density matrix, 
\begin{align}
\hat{\rho}'_{\rm RBM} = \frac{\hat{\rho}_{\rm RBM} + \hat{\rho}_{\rm RBM}^{\dagger}}{2},    
\end{align}
which assures the physical observables to be real-valued.


We update the parameters given in Eq.~(\ref{eqn:def_RBM}) so as to approximate the stationary state using the VMC sampling in the vector representation over the probability distribution 
\begin{align}
p(\sigma, \tau) = \frac{|\langle \braket{\sigma, \tau | \rho_{\rm RBM}}\rangle |^2 }{\langle\braket{\rho_{\rm RBM}|\rho_{\rm RBM}}\rangle}.
\end{align}
In the following, the number of the 
\textcolor{black}{sampled spin configurations at each step of optimization} is denoted as $N_s$.
\textcolor{black}{The parameters in the NSS ansatz are updated using} the stochastic reconfiguration method \cite{sorella_2001}, which is also known as the natural gradient method \cite{amari_1992, amari_1998}.
This optimization, being equivalent to the sufficiently long imaginary-time evolution in the truncated Hilbert space spanned by variational ansatz \cite{nomura_2017},  successfully avoids the local minima and converges to the desired state. Such an update step is repeated for $N_{\rm it}$ times until the cost function reaches the order of $10^{-3}$ or less.

\section{Model and Result}\label{sec:model_result}
In this section, we first demonstrate that our NSS based on the RBM  is capable of 
\textcolor{black}{simulating a state with large complexity in the sense of the operator space entanglement entropy, which is defined as the entanglement entropy of the mixed state in the vector representation.}
We then  verify our NSS method by applying it to three models that are in principle experimentally realizable  using cold atoms or trapped ions~\cite{lee_2013}: the transverse-field Ising models in 1D and 2D as well as the XYZ model in 1D.

\subsection{Random-valued NSS}\label{subsec:random_dm}


The RBMs for pure states are known to be capable of expressing quantum states with large entanglement efficiently. 
Concretely, Ref.~\cite{deng_2017} has shown that the maximally entangled states can be expressed using only $O(L)$ hidden spins, where $L$ is the total number of spins in the system.

Similarly, we argue that the NSS ansatz given as Eq.~\eqref{eqn:def_RBM} efficiently expresses density matrices with large operator space entanglement, 
\textcolor{black}{namely} the entanglement entropy of the density matrix in the vector representation~\cite{prosen_2007,pizorn_2009}.
\textcolor{black}{Concrete definition of the operator space entanglement entropy throughout this paper is given as follows. 
Let a mixed state in the vector representation, $\ket{\rho\rangle}$, be a pure state on the doubled Hilbert space spanned by $L$ physical and $L$ fictitious spins. 
After choosing $[L/2]$ physical spins and corresponding $[L/2]$ fictitious spins to form a subsystem $\mathcal{S}$, we  compute the entanglement entropy of $\mathrm{Tr}_{\mathcal{\bar{\mathcal{S}}}}[\ket{\rho\rangle}\bra{\langle\rho}]$, where $\bar{\mathcal{S}}$ is the complement of $\mathcal{S}$. Here, $[x]$ denotes the largest integer that does not exceed $x$.
}

\begin{figure}[t]
\begin{center}
\begin{tabular}{c}
  \begin{minipage}{0.97\hsize}
    \begin{center}
     \resizebox{0.95\hsize}{!}{\includegraphics{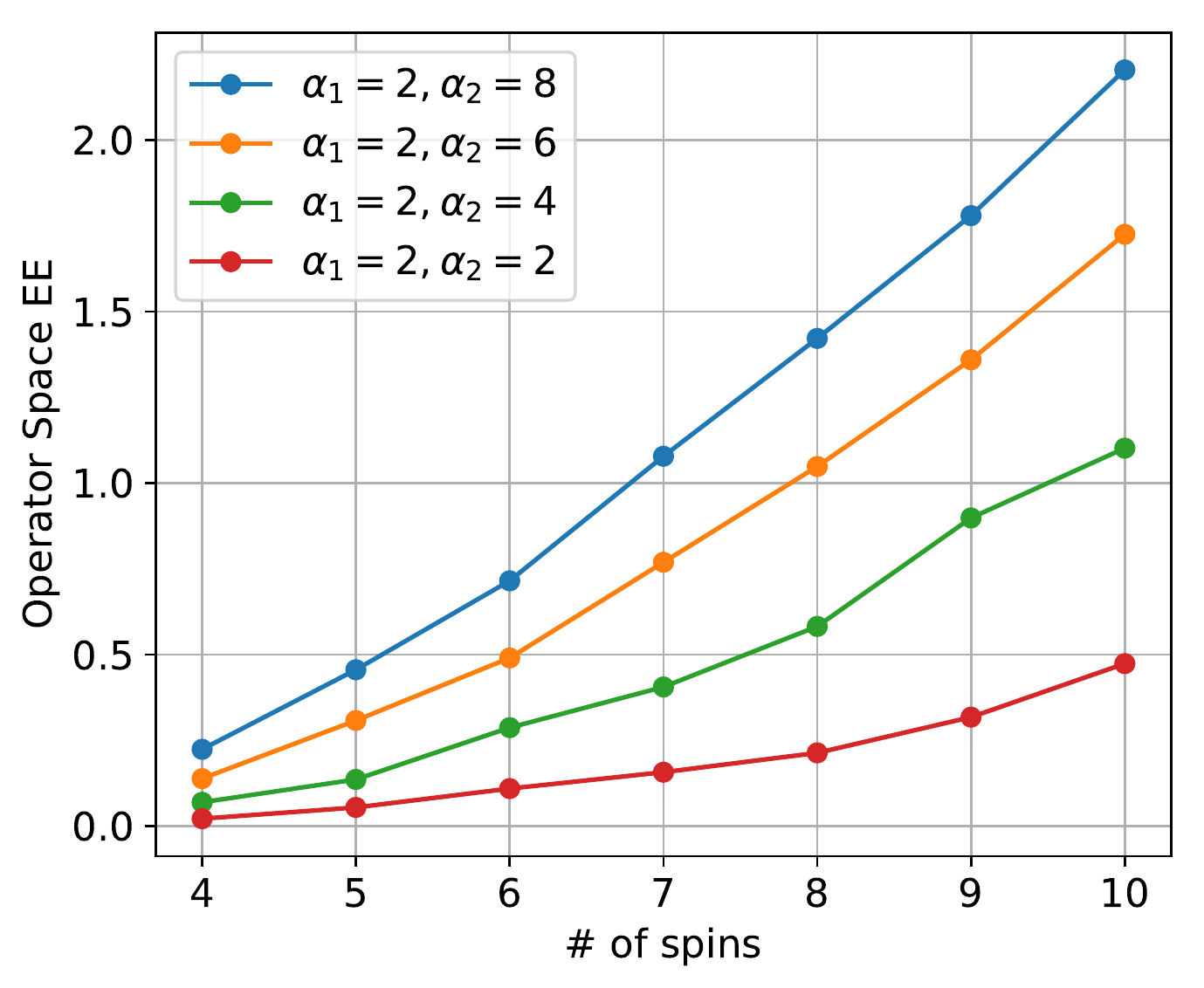}}

   \end{center}
  \end{minipage}
\end{tabular}
\end{center}
\caption{\label{fig:random_NSS} (Color Online) \textcolor{black}{The volume-law scaling of the operator space entanglement entropy in the random-valued NSS ansatz. Here, the number ratios of the hidden spins are taken as $(\alpha_1, \alpha_2) = (2, 8), (2, 6), (2, 4), (2,2)$. We do not observe any evident difference between finite $\alpha_1$, and hence we fix as $\alpha_1 = 2$. The amplitudes of the random parameters are set as $r = 0.05, 0.003, 0.1, 0.1$ for $W_{ij}, W_{ik}, a_i, b_k$, respectively. 
Note that each points show the averaged value over 10 independently generated random states.
}
}
\end{figure}

\textcolor{black}{
To demonstrate our argument, we show that generic NSS ansatz with random parameters exhibits volume-law scaling of the operator space entanglement entropy.
Shown in Fig.~\ref{fig:random_NSS} is the sistem size dependence of the operator space entanglement entropy in random-valued NSS ansatz characterized by several different parameters (see the next paragraph). 
The quantum entanglement in the operator space seems to increase along the number of spins, which demonstrates the volume-law scaling.
We thus argue that the large operator space entanglement entropy is not necessarily an obstacle for reliable simulations for our NSS,
in contrast with methods based on the tensor network ansatz such as the MPO algorithm.
As a caveat, we note that not all volume-law states can be expressed efficiently by NSS as is discussed in Appendix \ref{sec:random_fit}.}

\textcolor{black}{
The detailed calculation for random-valued RBM is done as follows, which is necessarily to justify the positive-semidefiniteness of the state~\cite{torlai_2018}. 
A subset of hidden spins, which are labeled by $j$ with the number ratio of spins denoted as $\alpha_1$, are connected to both physical and fictitious spins. The interactions between the $i$-th physical (fictitious) spins, denoted as $W_{ij}(\bar{W}_{ij})$ are required to satisfy $W_{ij} = \bar{W}_{ij}^*$ and also the magnetic field to be $b_j=0$. 
The rest of the hidden spins are connected to either only physical or fictitious spins. Denoting the labels for such hidden spins as $k$ and $\bar{k}$, the parameters  obey $W_{ik} = \bar{W}_{i\bar{k}}^*$ and $b_{k} = {b}_{\bar{k}}^*$ while the other parameters are zero. The number ratio of spins for such hidden spins are given by $\alpha_2$ each, and hence the total is given as $\alpha = \alpha_1 + 2\alpha_2$.
Under such conditions, both the real and imaginary parts of the parameters are drawn randomly from a section $[-r, r]$.
}

\subsection{Transverse-field Ising model in one dimension}\label{subsec:1dTFIM}
We now discuss the validity of our NSS ansatz for the concrete open quantum many-body systems.
We first consider the stationary state of a 1D transverse-field Ising model with the length $L$ under the periodic boundary condition. The Hamiltonian and the jump operators are given as
\begin{align}\label{eqn:1DTFIM}
 \hat{H} &= \frac{V}{4} \sum_{i=0}^{L-1} \hat{\sigma}_i^z \hat{\sigma}_{i+1}^z
 + \frac{g}{2} \sum_{i=0}^{L-1} \hat{\sigma}_i^x, \\
 \hat{\Gamma}_i &= \hat{\sigma}_i^-, \ \ \ \gamma_i = \gamma,
\end{align}
where $\hat{\sigma}_i^a\ (a = x,y,z)$ is the Pauli matrix that acts on the $i$-th site, $V$ is strength of the nearest-neighbor interaction, $g$ is amplitude of the  transverse field along the $x$-axis, and $\gamma$ gives the magnitude of the homogeneous dissipations.
To take advantage of the periodic boundary condition, i.e., $\hat{\sigma}_{L} = \hat{\sigma}_0$, we impose translation symmetry on the NSS ansatz.

\begin{figure}[t]
\begin{center}
\begin{tabular}{c}
  \begin{minipage}{0.97\hsize}
    \begin{center}
     \resizebox{0.99\hsize}{!}{\includegraphics{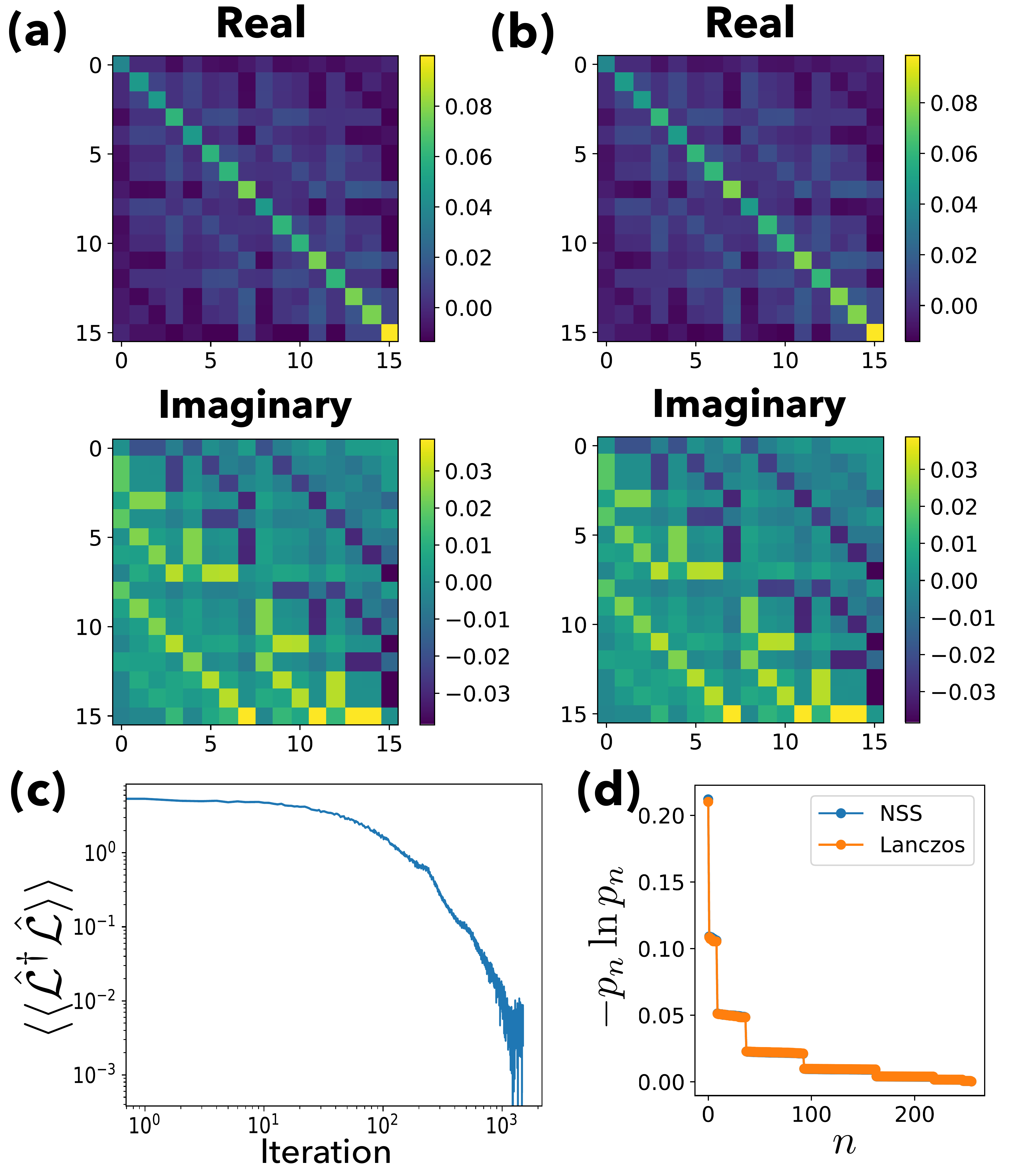}}
   \end{center}
  \end{minipage}
\end{tabular}
\end{center}
\caption{\label{fig:1dTFIM_densitymat_L8} (Color Online) (a) The real and imaginary parts of the stationary-state density matrix of the  1D transverse-field Ising model with dissipations in Eq.~\eqref{eqn:1DTFIM} obtained by the Lanczos method. (b) The real and imaginary parts of the stationary-state density matrix obtained by the NSS ansatz. 
The fidelity $F$ between the stationary states obtained by the Lanczos method and the NSS is over 0.999.
(c) Optimization of the cost function $\langle\braket{\hat{\Lind}^{\dagger}\hat{\Lind}}\rangle$. 
The optimization works well and $\langle\braket{\hat{\Lind}^{\dagger}\hat{\Lind}}\rangle$ reaches the order of $10^{-3}$.
(d) The entropy contribution $-p_n \ln p_n$, where $p_n$ is the $n$-th eigenvalue of the density matrix from the top. The blue and orange dots denote the results for the NSS ansatz and Lanczos method, respectively. The {relative} error for the total entropy is order of $10^{-3}$.
For all panels, we use $V= 0.3, g=1,$ and $\gamma = 0.5$, and the number ratio of the spins is $\alpha=1$. 
The sampling number per iteration is $N_s = 2000$, repeated for $N_\mathrm{it} = 1500$ iterations.
\textcolor{black}{The system sizes are given as $L=4$ for (a) and (b), while (c) and (d) are calculated for $L=8$.}
}
\end{figure}

As was introduced in Sec.~\ref{sec:NSS}, we optimize the expectation value $ \langle\braket{\hat{\Lind}^{\dagger} \hat{\Lind}}\rangle$ using the stochastic reconfiguration method. 
Figure.~\ref{fig:1dTFIM_densitymat_L8} shows the comparison of stationary-state density matrices obtained by the Lanczos method\textcolor{black}{, which efficiently approximates a subset of eigenvectors and eigenvalues of a sparse matrix \cite{arpack_1998}}, and NSS ansatz 
\textcolor{black}{with the number ratio of the spins taken as $\alpha=1$.}
Here, the model parameters are taken as $V=0.3, g = 1,$ and $\gamma = 0.5$, which results in a stationary state with the volume-law entropy. 
\textcolor{black}{ Figure~\ref{fig:1dTFIM_densitymat_L8}(a)(b) visually illustrates that the approximation of the state with the NSS well represents the stationary state calculated by the Lanczos method.
The accuracy of the stationary state is also confirmed  quantitatively via the calculation of the fidelity. 
The fidelity between $\hat{\rho}_1$ and $\hat{\rho}_2$, which is exclusively considered as the stationary-state density matrices obtained by the NSS optimization and Lanczos method in practice, is defined as \cite{jozsa_1994}
\begin{eqnarray}
F(\hat{\rho}_1, \hat{\rho}_2) = \left( \mathrm{Tr}\sqrt{\sqrt{\hat{\rho}_1} \hat{\rho}_2 \sqrt{\hat{\rho}_1}} \right)^2.
\end{eqnarray}
This corresponds to the largest fidelity between any two purifications of the density matrices.
For the current case, we find the fidelity to satisfy $F>0.999$.}
We also observe in Fig. \ref{fig:1dTFIM_densitymat_L8}(c) that the expectation value  $\langle\braket{\hat{\Lind}^{\dagger} \hat{\Lind}}\rangle$\textcolor{black}{, which gives measure of the approximation~\footnote{We can show that $|\braket{\braket{{\rho}_\mathrm{SS}|{\rho}_\mathrm{RBM}}}|^2\geq 1-\frac{\langle\braket{\hat{\Lind}^{\dagger} \hat{\Lind}}\rangle}{\Delta}$, where $\Delta$ is the spectral gap of $\hat{\Lind}^{\dagger} \hat{\Lind}$}, }
is nicely optimized and reaches the order of $10^{-3}$. Accordingly, the physical quantities are in good agreement with the exact results. 
For example, the entropy contribution for each eigenvalue of the density matrix, i.e., $-p_n \ln p_n$ for the $n$-th eigenvalue $p_n$, is remarkably accurate (see Fig.~\ref{fig:1dTFIM_densitymat_L8}(d)), such that the relative error of the total entropy is the order of $10^{-3}$. 

As is the case with other  VMC calculations, it must be noted that both  numerical cost and required memory for optimizing the NSS ansatz is much suppressed compared to methods that deal with the whole Hilbert space. In particular, the wall time for the NSS and the Lanczos methods are compared in
\textcolor{black}{Appendix~\ref{sec:walltime}.}

\subsection{Transverse-field Ising model in two dimension}\label{subsec:2dTFIM}
We next optimize the NSS ansatz for the 2D transverse-field Ising model on the square lattice with system size $L_x$ and $L_y$ along the $x$- and $y$-axes, respectively. We again take the periodic boundary condition.
The Hamiltonian and the jump operators are given as~\cite{jin_2018}
\begin{eqnarray}\label{eq:2Dising}
 \hat{H} &=& \frac{V}{4} \sum_{\braket{i,i'}} \hat{\sigma}_i^z \hat{\sigma}_{i'}^z
 + \frac{g}{2} \sum_i \hat{\sigma}_i^x, \\
 \hat{\Gamma}_i &=& \hat{\sigma}_i^-, \ \ \ \gamma_i = \gamma,
\end{eqnarray}
where the summation in the first term of $\hat{H}$ is taken over the edges connecting the neighboring sites, which are denoted as $i$ and $i'$. 

\begin{figure}[t]
\begin{center}
\begin{tabular}{c}
  \begin{minipage}{0.97\hsize}
    \begin{center}
     \resizebox{0.95\hsize}{!}{\includegraphics{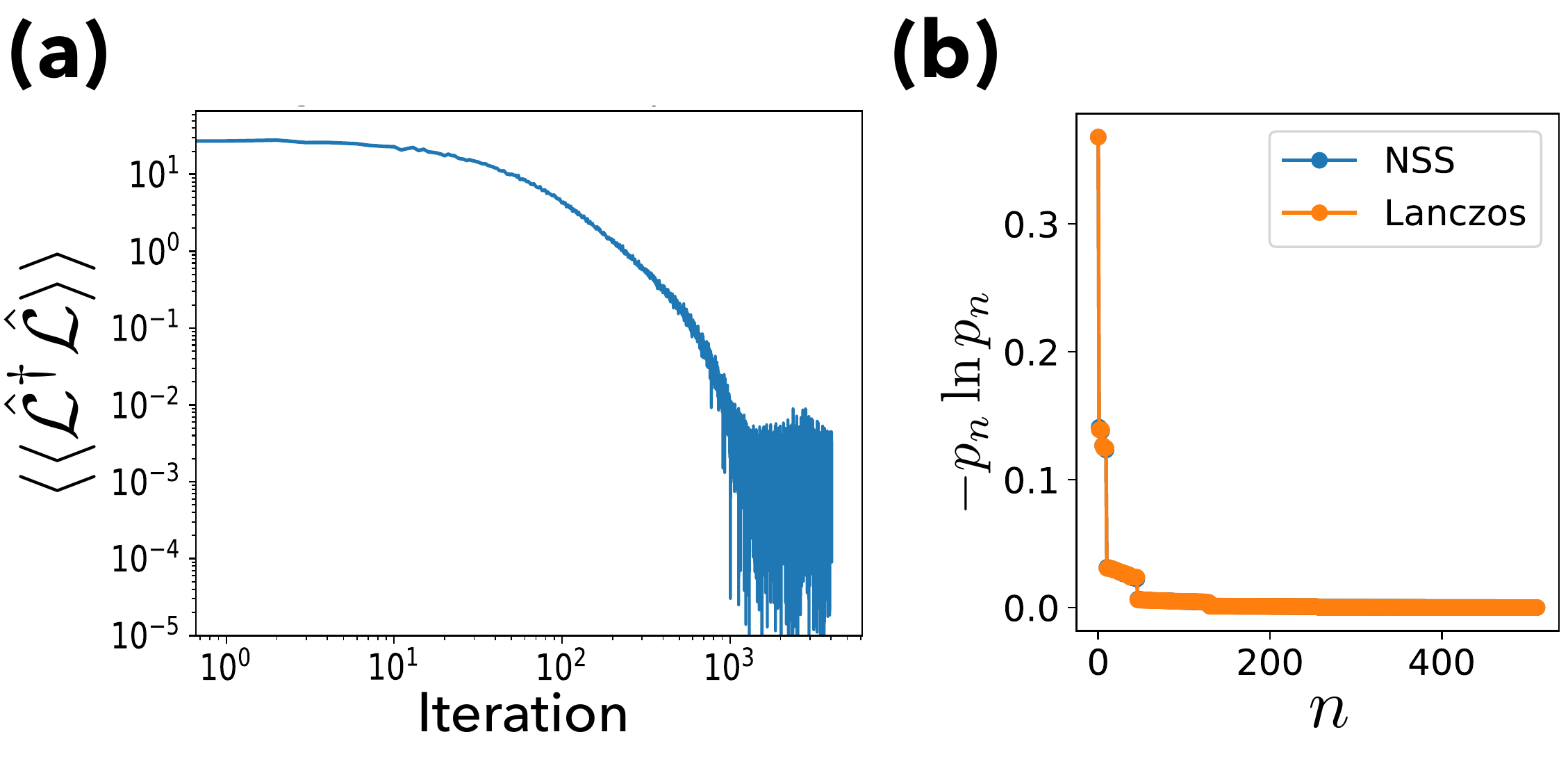}}

   \end{center}
  \end{minipage}
\end{tabular}
\end{center}
\caption{\label{fig:2dTFIM_eigval_3x3} (Color Online) (a) Optimization of the cost function $\braket{\braket{\Lind^{\dagger}\Lind}}$ for the 2D transverse-field Ising model in Eq.~\eqref{eq:2Dising}.
(b) The entropy contribution $-p_n \ln p_n$ for $n$-th eigenvalue. The relative error of the total entropy is order of $10^{-5}$.
We use the parameters $L_x =  L_y = 3, V = 0.3, g = 1$, and $\gamma = 1$. The number ratio  of the spins is $\alpha = 4$, and the resulting fidelity is $F = 0.9996$.
The sampling number per iteration is $N_s = 2000$, repeated for $N_\mathrm{it} = 4000$ iterations.
}
\end{figure}

The cost function in Figure~\ref{fig:2dTFIM_eigval_3x3}(a) shows that our optimization works well even for the 2D case.
Indeed, as shown in Fig.~\ref{fig:2dTFIM_eigval_3x3}(b), the NSS simulates the entropy contribution for each eigenvalue of the stationary state with high accuracy.
This result strengthens the expectation that our NSS ansatz does not suffer from high dimensionality, which can cause problems for the MPO ansatz.

\subsection{XYZ model in one dimension}\label{subsec:1dXYZ}
Finally, we investigate the 1D XYZ model, in which the dissipations are known to invoke dramatic change of the phase diagram compared with the closed system \cite{jiasen_2016}.
The model is defined as
\begin{eqnarray}\label{eqn:1DXYZ}
 \hat{H} &=& \sum_{i=0}^{L-1} J_x \hat{\sigma}_i^x \hat{\sigma}_{i+1}^x 
 + J_y \hat{\sigma}_i^y \hat{\sigma}_{i+1}^y
 + J_z \hat{\sigma}_i^z \hat{\sigma}_{i+1}^z \\
 \hat{\Gamma}_i &=& \hat{\sigma}_i^-, \ \ \ \gamma_i = \gamma,
\end{eqnarray}
where $J_a$ denotes the interaction for $a\ (a=x,y,z)$ component of the spin.
We particularly take $J_x = 0.9, J_y = 0.4, J_z = 1,$ and $\gamma=1$ with the periodic boundary condition, at which the finite system shows remnants of the phase transition predicted by the mean-field approximation~\cite{weimer_2015}.

\begin{figure}[t]
\begin{center}
\begin{tabular}{c}
  \begin{minipage}{0.6\hsize}
    \begin{center}
     \resizebox{0.95\hsize}{!}{\includegraphics{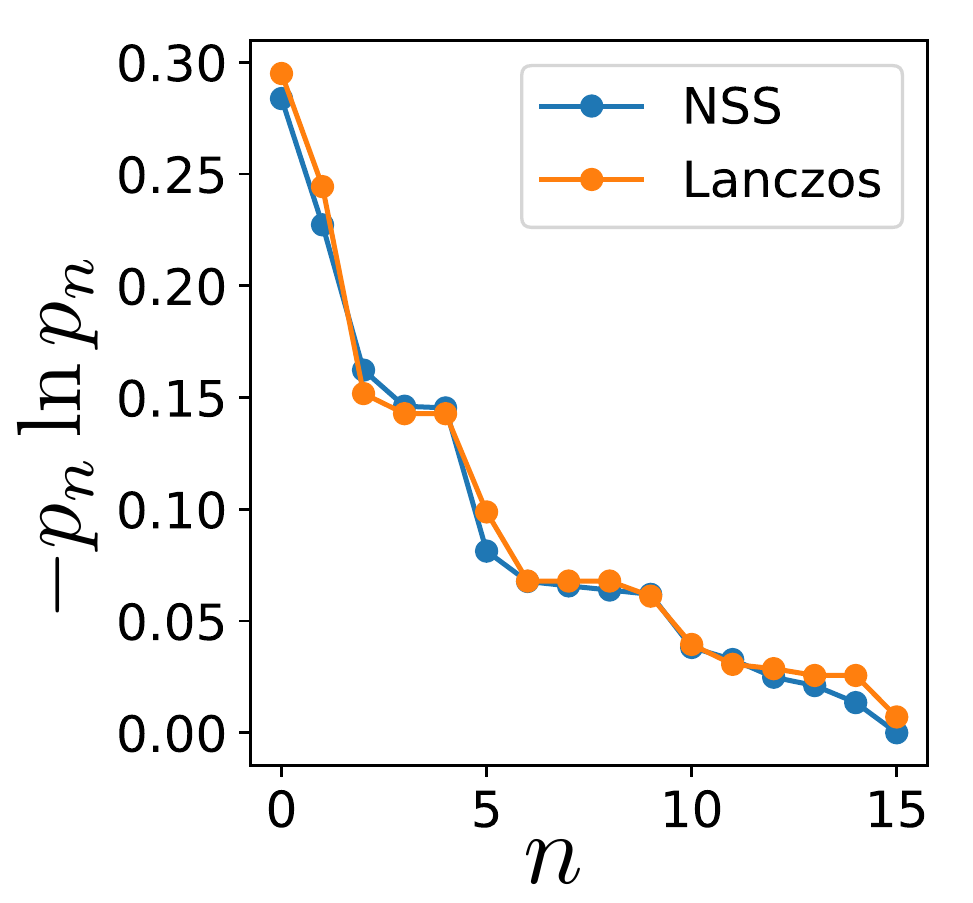}}

   \end{center}
  \end{minipage}
\end{tabular}
\end{center}
\caption{\label{fig:1dXYZ_eigval_L4} (Color Online) The entropy contribution $-p_n \ln p_n$  for the $n$-th eigenvalue of the stationary-state density matrix in 1D XYZ model in Eq.~\eqref{eqn:1DXYZ}. The data for the NSS (blue) and the exact diagonalization (orange) agree well with each other. 
The fidelity of the NSS is over 0.998 and the relative error of the total entropy is order of $10^{-2}$.
The parameters of the model is taken as $J_x = 0.9, J_y =0.4, J_z = 1.0, L = 4,$ and  $\gamma = 1$. The number ratio of the spins is taken as $\alpha = 8$ and the sampling number per iteration is $N_s = 8000$, repeated for $N_\mathrm{it} = 4500$ iterations.
}
\end{figure}

Shown in Fig.~\ref{fig:1dXYZ_eigval_L4} is the comparison of the translationally symmetric NSS ansatz and the Lanczos method regarding the  entropy  contribution for each eigenvalue.
Even though our choice of parameters leads to the  non-simple stationary state of our small systems (as indicated from the peak of the structure factor~\cite{jiasen_2016}), the NSS describes the exact results well.

\section{Conclusions and outlooks\label{sec:conclusions}}
We have proposed that the neural quantum states are suited for expressing the stationary states of open quantum many-body systems.
By mapping the original stationary-state search problem of the Lindblad equation to the zero-energy ground state search problem of an appropriate Hermitian operator, we solve it with a variational ansatz based on the RBM, or the neural stationary state (NSS).

We have confirmed that the NSS can even 
We have then demonstrated that our NSS ansatz is capable of expressing the stationary states of the dissipative one- and two-dimensional transverse-field Ising models and one-dimensional XYZ model. 

While the aim of our work is to make a first attempt to show the adequacy of the NSS for  open quantum many-body systems including highly entangled states and two-dimensional states, we leave several intriguing questions as future works.
One naive question is whether our ansatz can simulate larger system sizes, in which other methods suffer from expensive numerical cost.
Another important question is to clarify the versatility of open quantum many-body systems addressable by our method.
We expect that our ansatz performs well regardless of the  dimensionality,
as suggested in our calculations and the bipartite-graph structure of the RBM, which is free from the geometry of the underlying physical lattice.
It is also interesting whether our method works for various long-range interacting systems (such as the Haldane-Shastry model~\cite{deng_2017}) with dissipations, whose mixed stationary states can be highly entangled.

\textit{Note added.-}
After completion of our work, we became aware of 
\textcolor{black}{some related works. Refs.~\cite{carleo_2019,nagy_2019} discussed the time evolution and stationary states of open quantum many-body systems by using the complex RBM and Ref.~\cite{vicentini_2019} studied the approximation of the stationary states by the RBM ansatz.
}

\begin{acknowledgments}
We are grateful to Masahito Ueda for reading our manuscripts with valuable discussions and helpful advice.
We also thank Hosho Katsura, Zongping Gong, and Shunsuke Furukawa for fruitful comments.
The numerical calculations were carried out with the help of NetKet~\cite{netket}, SciPy~\cite{scipy_2001}, QuTiP~\cite{qutip}, and ARPACK~\cite{arpack_1998}.
This work was supported by
KAKENHI Grant No. JP18H01145 and
a Grant-in-Aid for Scientific Research on Innovative Areas ``Topological Materials Science" (KAKENHI Grant No. JP15H05855)
from the Japan Society for the Promotion of Science.
N. Y. and R. H. were supported by Advanced Leading Graduate Course for Photon Science (ALPS) of Japan Society for the Promotion of Science (JSPS).
N. Y. was supported by JSPS KAKENHI Grant-in-Aid for JSPS fellows Grant No. JP17J00743.
R. H. was supported by JSPS KAKENHI Grant-in-Aid for JSPS fellows Grant No. JP17J03189).
\end{acknowledgments}

\appendix

\section{Approximating Random Density Matrices by NSS}\label{sec:random_fit}
\textcolor{black}{In this appendix, we randomly generate a density matrix and fit it by the NSS to see that the expressive power of the ansatz does not assure efficient representation of all volume-law states.
Here, random density matrices are generated as}
\begin{eqnarray}\label{eqn:randomdef}
\hat{\rho}=\frac{\hat{X}^2}{\mathrm{Tr}[\hat{X}^2]},
\end{eqnarray} 
\textcolor{black}{where $\hat{X}$ is sampled from the Gaussian unitary ensembles of random Hermitian matrices.
We have numerically checked that the operator space entanglement entropy defined as in the main text
exhibits a volume-law scaling, i.e., operator space entanglement entropy $\propto L$ for matrices with size $2^L\times 2^L$ (data not shown).}

\begin{figure}[t]
\begin{center}
\begin{tabular}{c}
  \begin{minipage}{0.97\hsize}
    \begin{center}
     \resizebox{0.95\hsize}{!}{\includegraphics{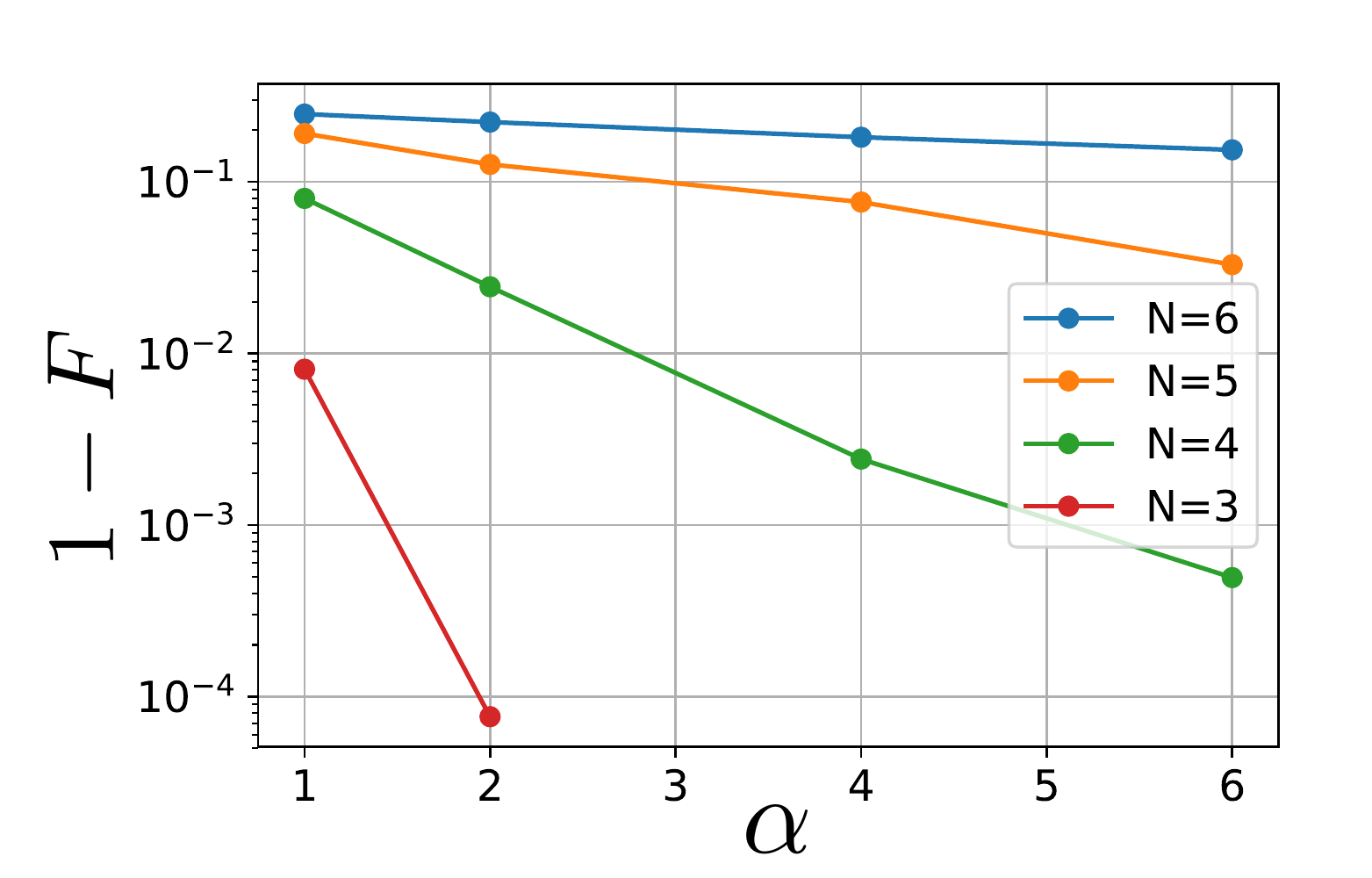}}

   \end{center}
  \end{minipage}
\end{tabular}
\end{center}
\caption{\label{fig:random_density}
\textcolor{black}{
(Color Online) The infidelity $1 - F$ of the NSS fitted to a random density matrix generated by Eq.~\eqref{eqn:randomdef}.
 While the NSS with larger $\alpha$, or the number ratio of the spins,
better approximates the random density matrices, the number of parameters and accordingly the numerical cost required to reach some fixed fidelity increase rapidly.
}
}
\end{figure}

\textcolor{black}{
Figure~\ref{fig:random_density} shows that while a random density matrix generated following Eq.~\eqref{eqn:randomdef} can be approximated better by the NSS with larger $\alpha$, or the number ratio of the spins, the number of parameters and accordingly the numerical cost required to reach some fixed fidelity increase rapidly.
}

\section{Comparison of Computational Cost}\label{sec:walltime}
\begin{figure}[t]
\begin{center}
\begin{tabular}{c}
  \begin{minipage}{0.97\hsize}
    \begin{center}
     \resizebox{0.95\hsize}{!}{\includegraphics{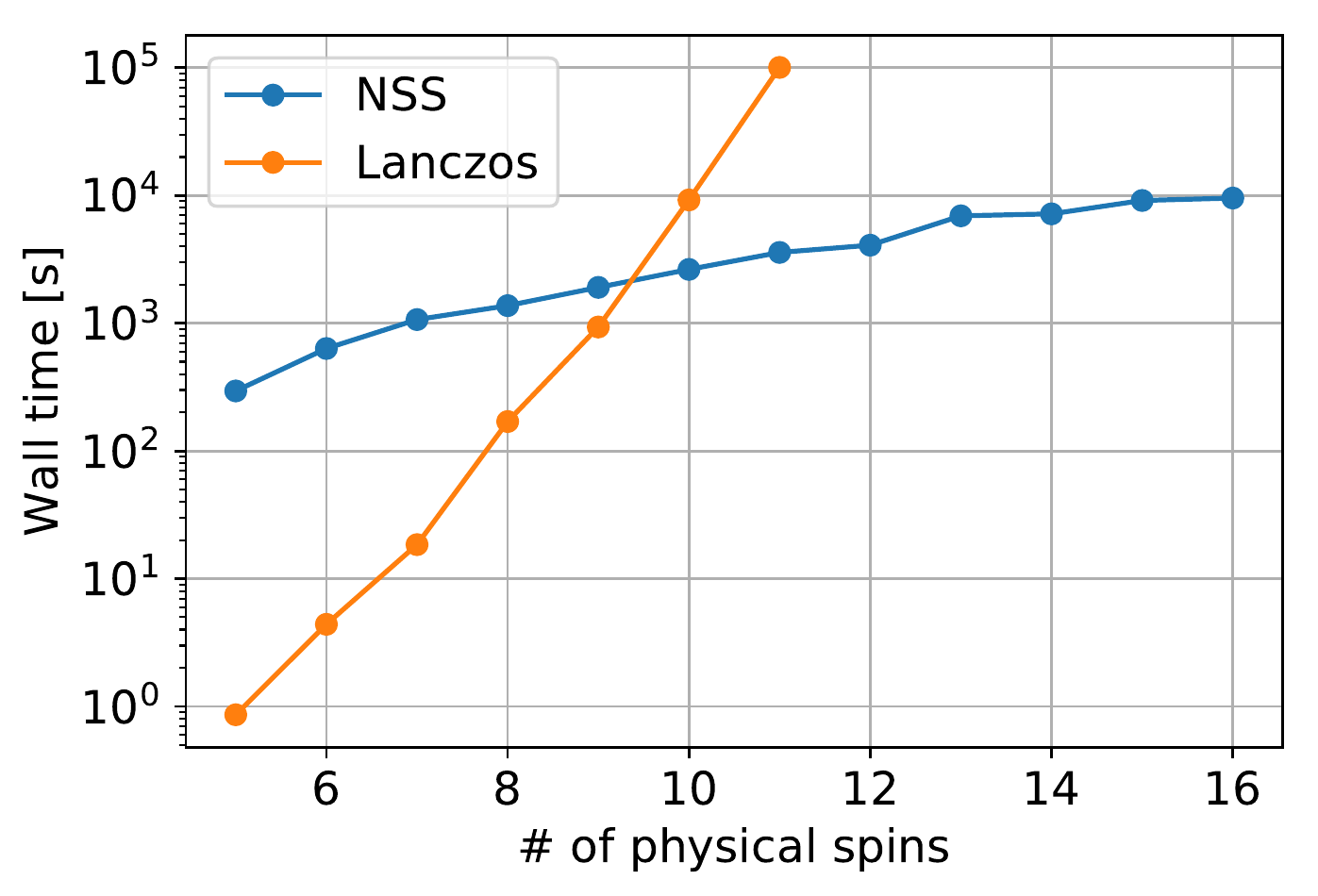}}

   \end{center}
  \end{minipage}
\end{tabular}
\end{center}
\caption{\label{fig:walltime}
\textcolor{black}{
(Color Online) 
 The wall times for computing the stationary state of the 1D transverse-field Ising model with $V=2, g = 1$, and $\gamma = 1$. The blue and orange dots are for the NSS ansatz optimization and the Lanczos method, respectively. Here, the number ratio of the spins is  $\alpha = 4$. 
 The NSS ansatz exhibits lower scaling as a function of the number of physical spins.
 The number of sampling is $N_s = 2000$ repeated for $N_{\rm it} = 1500$ iterations, which we find to be sufficient for the convergence of the VMC calculation.
 The computation for the NSS and Lanczos is executed on 8 cores on Intel(R) Core i7-6820HQ and 12 cores on Intel(R) Xeon(R) Silver 4110, respectively.}
 }
\end{figure}

\textcolor{black}{In Fig.~\ref{fig:walltime}, we show the scaling of the computational time for calculating the stationary states by optimization of the NSS and the Lanczos method. 
Our variational method exhibits only polynomial scaling which is, clearly, far more efficient than the exponential scaling in the Lanczos method.}

\bibliographystyle{apsrev4-1}
\bibliography{bibliography_openrbm.bib}

\begin{thebibliography}{72}%
\makeatletter
\providecommand \@ifxundefined [1]{%
 \@ifx{#1\undefined}
}%
\providecommand \@ifnum [1]{%
 \ifnum #1\expandafter \@firstoftwo
 \else \expandafter \@secondoftwo
 \fi
}%
\providecommand \@ifx [1]{%
 \ifx #1\expandafter \@firstoftwo
 \else \expandafter \@secondoftwo
 \fi
}%
\providecommand \natexlab [1]{#1}%
\providecommand \enquote  [1]{``#1''}%
\providecommand \bibnamefont  [1]{#1}%
\providecommand \bibfnamefont [1]{#1}%
\providecommand \citenamefont [1]{#1}%
\providecommand \href@noop [0]{\@secondoftwo}%
\providecommand \href [0]{\begingroup \@sanitize@url \@href}%
\providecommand \@href[1]{\@@startlink{#1}\@@href}%
\providecommand \@@href[1]{\endgroup#1\@@endlink}%
\providecommand \@sanitize@url [0]{\catcode `\\12\catcode `\$12\catcode
  `\&12\catcode `\#12\catcode `\^12\catcode `\_12\catcode `\%12\relax}%
\providecommand \@@startlink[1]{}%
\providecommand \@@endlink[0]{}%
\providecommand \url  [0]{\begingroup\@sanitize@url \@url }%
\providecommand \@url [1]{\endgroup\@href {#1}{\urlprefix }}%
\providecommand \urlprefix  [0]{URL }%
\providecommand \Eprint [0]{\href }%
\providecommand \doibase [0]{http://dx.doi.org/}%
\providecommand \selectlanguage [0]{\@gobble}%
\providecommand \bibinfo  [0]{\@secondoftwo}%
\providecommand \bibfield  [0]{\@secondoftwo}%
\providecommand \translation [1]{[#1]}%
\providecommand \BibitemOpen [0]{}%
\providecommand \bibitemStop [0]{}%
\providecommand \bibitemNoStop [0]{.\EOS\space}%
\providecommand \EOS [0]{\spacefactor3000\relax}%
\providecommand \BibitemShut  [1]{\csname bibitem#1\endcsname}%
\let\auto@bib@innerbib\@empty
\bibitem [{\citenamefont {Carrasquilla}\ and\ \citenamefont
  {Melko}(2017)}]{carrasquilla_2017}%
  \BibitemOpen
  \bibfield  {author} {\bibinfo {author} {\bibfnamefont {J.}~\bibnamefont
  {Carrasquilla}}\ and\ \bibinfo {author} {\bibfnamefont {R.~G.}\ \bibnamefont
  {Melko}},\ }\href {\doibase 10.1038/nphys4035} {\bibfield  {journal}
  {\bibinfo  {journal} {Nat. Phys.}\ }\textbf {\bibinfo {volume} {13}},\
  \bibinfo {pages} {431} (\bibinfo {year} {2017})}\BibitemShut {NoStop}%
\bibitem [{\citenamefont {Carleo}\ and\ \citenamefont
  {Troyer}(2017)}]{carleo_2017}%
  \BibitemOpen
  \bibfield  {author} {\bibinfo {author} {\bibfnamefont {G.}~\bibnamefont
  {Carleo}}\ and\ \bibinfo {author} {\bibfnamefont {M.}~\bibnamefont
  {Troyer}},\ }\href {\doibase 10.1126/science.aag2302} {\bibfield  {journal}
  {\bibinfo  {journal} {Science}\ }\textbf {\bibinfo {volume} {355}},\ \bibinfo
  {pages} {602} (\bibinfo {year} {2017})}\BibitemShut {NoStop}%
\bibitem [{\citenamefont {Wang}(2016)}]{leiwang_2016}%
  \BibitemOpen
  \bibfield  {author} {\bibinfo {author} {\bibfnamefont {L.}~\bibnamefont
  {Wang}},\ }\href {\doibase 10.1103/PhysRevB.94.195105} {\bibfield  {journal}
  {\bibinfo  {journal} {Phys. Rev. B}\ }\textbf {\bibinfo {volume} {94}},\
  \bibinfo {pages} {195105} (\bibinfo {year} {2016})}\BibitemShut {NoStop}%
\bibitem [{\citenamefont {Nieuwenburg}\ \emph {et~al.}(2017)\citenamefont
  {Nieuwenburg}, \citenamefont {Liu},\ and\ \citenamefont
  {Huber}}]{evert_2017}%
  \BibitemOpen
  \bibfield  {author} {\bibinfo {author} {\bibfnamefont {E.~P.~L.}\
  \bibnamefont {Nieuwenburg}}, \bibinfo {author} {\bibfnamefont {Y.-H.}\
  \bibnamefont {Liu}}, \ and\ \bibinfo {author} {\bibfnamefont {S.~D.}\
  \bibnamefont {Huber}},\ }\href@noop {} {\bibfield  {journal} {\bibinfo
  {journal} {Nat. Phys.}\ }\textbf {\bibinfo {volume} {13}},\ \bibinfo {pages}
  {435} (\bibinfo {year} {2017})}\BibitemShut {NoStop}%
\bibitem [{\citenamefont {Hu}\ \emph {et~al.}(2017)\citenamefont {Hu},
  \citenamefont {Singh},\ and\ \citenamefont {Scalettar}}]{hu_2017}%
  \BibitemOpen
  \bibfield  {author} {\bibinfo {author} {\bibfnamefont {W.}~\bibnamefont
  {Hu}}, \bibinfo {author} {\bibfnamefont {R.~R.~P.}\ \bibnamefont {Singh}}, \
  and\ \bibinfo {author} {\bibfnamefont {R.~T.}\ \bibnamefont {Scalettar}},\
  }\href {\doibase 10.1103/PhysRevE.95.062122} {\bibfield  {journal} {\bibinfo
  {journal} {Phys. Rev. E}\ }\textbf {\bibinfo {volume} {95}},\ \bibinfo
  {pages} {062122} (\bibinfo {year} {2017})}\BibitemShut {NoStop}%
\bibitem [{\citenamefont {Broecker}\ \emph {et~al.}(2017)\citenamefont
  {Broecker}, \citenamefont {Carrasquilla}, \citenamefont {Melko},\ and\
  \citenamefont {Trebst}}]{broecker_sign_2017}%
  \BibitemOpen
  \bibfield  {author} {\bibinfo {author} {\bibfnamefont {P.}~\bibnamefont
  {Broecker}}, \bibinfo {author} {\bibfnamefont {J.}~\bibnamefont
  {Carrasquilla}}, \bibinfo {author} {\bibfnamefont {R.~G.}\ \bibnamefont
  {Melko}}, \ and\ \bibinfo {author} {\bibfnamefont {S.}~\bibnamefont
  {Trebst}},\ }\href@noop {} {\bibfield  {journal} {\bibinfo  {journal}
  {Scientific Reports}\ }\textbf {\bibinfo {volume} {7}} (\bibinfo {year}
  {2017})}\BibitemShut {NoStop}%
\bibitem [{\citenamefont {Schindler}\ \emph {et~al.}(2017)\citenamefont
  {Schindler}, \citenamefont {Regnault},\ and\ \citenamefont
  {Neupert}}]{schindler_2017}%
  \BibitemOpen
  \bibfield  {author} {\bibinfo {author} {\bibfnamefont {F.}~\bibnamefont
  {Schindler}}, \bibinfo {author} {\bibfnamefont {N.}~\bibnamefont {Regnault}},
  \ and\ \bibinfo {author} {\bibfnamefont {T.}~\bibnamefont {Neupert}},\ }\href
  {\doibase 10.1103/PhysRevB.95.245134} {\bibfield  {journal} {\bibinfo
  {journal} {Phys. Rev. B}\ }\textbf {\bibinfo {volume} {95}},\ \bibinfo
  {pages} {245134} (\bibinfo {year} {2017})}\BibitemShut {NoStop}%
\bibitem [{\citenamefont {Liu}\ and\ \citenamefont {van
  Nieuwenburg}(2018)}]{liu_2018}%
  \BibitemOpen
  \bibfield  {author} {\bibinfo {author} {\bibfnamefont {Y.-H.}\ \bibnamefont
  {Liu}}\ and\ \bibinfo {author} {\bibfnamefont {E.~P.~L.}\ \bibnamefont {van
  Nieuwenburg}},\ }\href {\doibase 10.1103/PhysRevLett.120.176401} {\bibfield
  {journal} {\bibinfo  {journal} {Phys. Rev. Lett.}\ }\textbf {\bibinfo
  {volume} {120}},\ \bibinfo {pages} {176401} (\bibinfo {year}
  {2018})}\BibitemShut {NoStop}%
\bibitem [{\citenamefont {Zhang}\ \emph {et~al.}(2018)\citenamefont {Zhang},
  \citenamefont {Shen},\ and\ \citenamefont {Zhai}}]{zhang_shen_2018}%
  \BibitemOpen
  \bibfield  {author} {\bibinfo {author} {\bibfnamefont {P.}~\bibnamefont
  {Zhang}}, \bibinfo {author} {\bibfnamefont {H.}~\bibnamefont {Shen}}, \ and\
  \bibinfo {author} {\bibfnamefont {H.}~\bibnamefont {Zhai}},\ }\href@noop {}
  {\bibfield  {journal} {\bibinfo  {journal} {Phys. Rev. Lett.}\ }\textbf
  {\bibinfo {volume} {120}},\ \bibinfo {pages} {066401} (\bibinfo {year}
  {2018})}\BibitemShut {NoStop}%
\bibitem [{\citenamefont {Yoshioka}\ \emph
  {et~al.}(2018{\natexlab{a}})\citenamefont {Yoshioka}, \citenamefont {Akagi},\
  and\ \citenamefont {Katsura}}]{yoshioka_18}%
  \BibitemOpen
  \bibfield  {author} {\bibinfo {author} {\bibfnamefont {N.}~\bibnamefont
  {Yoshioka}}, \bibinfo {author} {\bibfnamefont {Y.}~\bibnamefont {Akagi}}, \
  and\ \bibinfo {author} {\bibfnamefont {H.}~\bibnamefont {Katsura}},\ }\href
  {\doibase 10.1103/PhysRevB.97.205110} {\bibfield  {journal} {\bibinfo
  {journal} {Phys. Rev. B}\ }\textbf {\bibinfo {volume} {97}},\ \bibinfo
  {pages} {205110} (\bibinfo {year} {2018}{\natexlab{a}})}\BibitemShut
  {NoStop}%
\bibitem [{\citenamefont {Torlai}\ and\ \citenamefont
  {Melko}(2016)}]{torlai_2016}%
  \BibitemOpen
  \bibfield  {author} {\bibinfo {author} {\bibfnamefont {G.}~\bibnamefont
  {Torlai}}\ and\ \bibinfo {author} {\bibfnamefont {R.~G.}\ \bibnamefont
  {Melko}},\ }\href {\doibase 10.1103/PhysRevB.94.165134} {\bibfield  {journal}
  {\bibinfo  {journal} {Phys. Rev. B}\ }\textbf {\bibinfo {volume} {94}},\
  \bibinfo {pages} {165134} (\bibinfo {year} {2016})}\BibitemShut {NoStop}%
\bibitem [{\citenamefont {Liu}\ \emph {et~al.}(2017)\citenamefont {Liu},
  \citenamefont {Qi}, \citenamefont {Meng},\ and\ \citenamefont
  {Fu}}]{j_liu_2017}%
  \BibitemOpen
  \bibfield  {author} {\bibinfo {author} {\bibfnamefont {J.}~\bibnamefont
  {Liu}}, \bibinfo {author} {\bibfnamefont {Y.}~\bibnamefont {Qi}}, \bibinfo
  {author} {\bibfnamefont {Z.~Y.}\ \bibnamefont {Meng}}, \ and\ \bibinfo
  {author} {\bibfnamefont {L.}~\bibnamefont {Fu}},\ }\href {\doibase
  10.1103/PhysRevB.95.041101} {\bibfield  {journal} {\bibinfo  {journal} {Phys.
  Rev. B}\ }\textbf {\bibinfo {volume} {95}},\ \bibinfo {pages} {041101}
  (\bibinfo {year} {2017})}\BibitemShut {NoStop}%
\bibitem [{\citenamefont {Huang}\ and\ \citenamefont {Wang}(2017)}]{h_li_2017}%
  \BibitemOpen
  \bibfield  {author} {\bibinfo {author} {\bibfnamefont {L.}~\bibnamefont
  {Huang}}\ and\ \bibinfo {author} {\bibfnamefont {L.}~\bibnamefont {Wang}},\
  }\href {\doibase 10.1103/PhysRevB.95.035105} {\bibfield  {journal} {\bibinfo
  {journal} {Phys. Rev. B}\ }\textbf {\bibinfo {volume} {95}},\ \bibinfo
  {pages} {035105} (\bibinfo {year} {2017})}\BibitemShut {NoStop}%
\bibitem [{\citenamefont {Wang}(2017)}]{w_lei_2017}%
  \BibitemOpen
  \bibfield  {author} {\bibinfo {author} {\bibfnamefont {L.}~\bibnamefont
  {Wang}},\ }\href {\doibase 10.1103/PhysRevE.96.051301} {\bibfield  {journal}
  {\bibinfo  {journal} {Phys. Rev. E}\ }\textbf {\bibinfo {volume} {96}},\
  \bibinfo {pages} {051301} (\bibinfo {year} {2017})}\BibitemShut {NoStop}%
\bibitem [{\citenamefont {Shen}\ \emph {et~al.}(2018)\citenamefont {Shen},
  \citenamefont {Liu},\ and\ \citenamefont {Fu}}]{huitao_2018}%
  \BibitemOpen
  \bibfield  {author} {\bibinfo {author} {\bibfnamefont {H.}~\bibnamefont
  {Shen}}, \bibinfo {author} {\bibfnamefont {J.}~\bibnamefont {Liu}}, \ and\
  \bibinfo {author} {\bibfnamefont {L.}~\bibnamefont {Fu}},\ }\href {\doibase
  10.1103/PhysRevB.97.205140} {\bibfield  {journal} {\bibinfo  {journal} {Phys.
  Rev. B}\ }\textbf {\bibinfo {volume} {97}},\ \bibinfo {pages} {205140}
  (\bibinfo {year} {2018})}\BibitemShut {NoStop}%
\bibitem [{\citenamefont {Yoshioka}\ \emph
  {et~al.}(2018{\natexlab{b}})\citenamefont {Yoshioka}, \citenamefont {Akagi},\
  and\ \citenamefont {Katsura}}]{yoshioka_18_2}%
  \BibitemOpen
  \bibfield  {author} {\bibinfo {author} {\bibfnamefont {N.}~\bibnamefont
  {Yoshioka}}, \bibinfo {author} {\bibfnamefont {Y.}~\bibnamefont {Akagi}}, \
  and\ \bibinfo {author} {\bibfnamefont {H.}~\bibnamefont {Katsura}},\
  }\href@noop {} {\bibfield  {journal} {\bibinfo  {journal} {arXiv:1812.0526}\
  } (\bibinfo {year} {2018}{\natexlab{b}})}\BibitemShut {NoStop}%
\bibitem [{\citenamefont {Deng}\ \emph
  {et~al.}(2017{\natexlab{a}})\citenamefont {Deng}, \citenamefont {Li},\ and\
  \citenamefont {Das~Sarma}}]{deng_2017}%
  \BibitemOpen
  \bibfield  {author} {\bibinfo {author} {\bibfnamefont {D.-L.}\ \bibnamefont
  {Deng}}, \bibinfo {author} {\bibfnamefont {X.}~\bibnamefont {Li}}, \ and\
  \bibinfo {author} {\bibfnamefont {S.}~\bibnamefont {Das~Sarma}},\ }\href
  {\doibase 10.1103/PhysRevX.7.021021} {\bibfield  {journal} {\bibinfo
  {journal} {Phys. Rev. X}\ }\textbf {\bibinfo {volume} {7}},\ \bibinfo {pages}
  {021021} (\bibinfo {year} {2017}{\natexlab{a}})}\BibitemShut {NoStop}%
\bibitem [{\citenamefont {Deng}\ \emph
  {et~al.}(2017{\natexlab{b}})\citenamefont {Deng}, \citenamefont {Li},\ and\
  \citenamefont {Das~Sarma}}]{deng_2017_prb}%
  \BibitemOpen
  \bibfield  {author} {\bibinfo {author} {\bibfnamefont {D.-L.}\ \bibnamefont
  {Deng}}, \bibinfo {author} {\bibfnamefont {X.}~\bibnamefont {Li}}, \ and\
  \bibinfo {author} {\bibfnamefont {S.}~\bibnamefont {Das~Sarma}},\ }\href
  {\doibase 10.1103/PhysRevB.96.195145} {\bibfield  {journal} {\bibinfo
  {journal} {Phys. Rev. B}\ }\textbf {\bibinfo {volume} {96}},\ \bibinfo
  {pages} {195145} (\bibinfo {year} {2017}{\natexlab{b}})}\BibitemShut
  {NoStop}%
\bibitem [{\citenamefont {Gao}\ and\ \citenamefont {Duan}(2017)}]{gao_2017}%
  \BibitemOpen
  \bibfield  {author} {\bibinfo {author} {\bibfnamefont {X.}~\bibnamefont
  {Gao}}\ and\ \bibinfo {author} {\bibfnamefont {L.-M.}\ \bibnamefont {Duan}},\
  }\href {\doibase 10.1038/s41467-017-00705-2} {\bibfield  {journal} {\bibinfo
  {journal} {Nature Communications}\ }\textbf {\bibinfo {volume} {8}},\
  \bibinfo {pages} {662} (\bibinfo {year} {2017})}\BibitemShut {NoStop}%
\bibitem [{\citenamefont {Nomura}\ \emph {et~al.}(2017)\citenamefont {Nomura},
  \citenamefont {Darmawan}, \citenamefont {Yamaji},\ and\ \citenamefont
  {Imada}}]{nomura_2017}%
  \BibitemOpen
  \bibfield  {author} {\bibinfo {author} {\bibfnamefont {Y.}~\bibnamefont
  {Nomura}}, \bibinfo {author} {\bibfnamefont {A.~S.}\ \bibnamefont
  {Darmawan}}, \bibinfo {author} {\bibfnamefont {Y.}~\bibnamefont {Yamaji}}, \
  and\ \bibinfo {author} {\bibfnamefont {M.}~\bibnamefont {Imada}},\ }\href
  {\doibase 10.1103/PhysRevB.96.205152} {\bibfield  {journal} {\bibinfo
  {journal} {Phys. Rev. B}\ }\textbf {\bibinfo {volume} {96}},\ \bibinfo
  {pages} {205152} (\bibinfo {year} {2017})}\BibitemShut {NoStop}%
\bibitem [{\citenamefont {Saito}(2017)}]{saito_2017}%
  \BibitemOpen
  \bibfield  {author} {\bibinfo {author} {\bibfnamefont {H.}~\bibnamefont
  {Saito}},\ }\href {\doibase 10.7566/JPSJ.86.093001} {\bibfield  {journal}
  {\bibinfo  {journal} {J. Phys. Soc. Jpn.}\ }\textbf {\bibinfo {volume}
  {86}},\ \bibinfo {pages} {093001} (\bibinfo {year} {2017})}\BibitemShut
  {NoStop}%
\bibitem [{\citenamefont {Carleo}\ \emph {et~al.}(2018)\citenamefont {Carleo},
  \citenamefont {Nomura},\ and\ \citenamefont {Imada}}]{carleo_2018}%
  \BibitemOpen
  \bibfield  {author} {\bibinfo {author} {\bibfnamefont {G.}~\bibnamefont
  {Carleo}}, \bibinfo {author} {\bibfnamefont {Y.}~\bibnamefont {Nomura}}, \
  and\ \bibinfo {author} {\bibfnamefont {M.}~\bibnamefont {Imada}},\
  }\href@noop {} {\bibfield  {journal} {\bibinfo  {journal} {Nature
  communications}\ }\textbf {\bibinfo {volume} {9}},\ \bibinfo {pages} {5322}
  (\bibinfo {year} {2018})}\BibitemShut {NoStop}%
\bibitem [{\citenamefont {Saito}\ and\ \citenamefont
  {Kato}(2018)}]{saito_2018}%
  \BibitemOpen
  \bibfield  {author} {\bibinfo {author} {\bibfnamefont {H.}~\bibnamefont
  {Saito}}\ and\ \bibinfo {author} {\bibfnamefont {M.}~\bibnamefont {Kato}},\
  }\href {\doibase 10.7566/JPSJ.87.014001} {\bibfield  {journal} {\bibinfo
  {journal} {J. Phys. Soc. Jpn.}\ }\textbf {\bibinfo {volume} {87}},\ \bibinfo
  {pages} {014001} (\bibinfo {year} {2018})}\BibitemShut {NoStop}%
\bibitem [{\citenamefont {Glasser}\ \emph {et~al.}(2018)\citenamefont
  {Glasser}, \citenamefont {Pancotti}, \citenamefont {August}, \citenamefont
  {Rodriguez},\ and\ \citenamefont {Cirac}}]{glasser_2018}%
  \BibitemOpen
  \bibfield  {author} {\bibinfo {author} {\bibfnamefont {I.}~\bibnamefont
  {Glasser}}, \bibinfo {author} {\bibfnamefont {N.}~\bibnamefont {Pancotti}},
  \bibinfo {author} {\bibfnamefont {M.}~\bibnamefont {August}}, \bibinfo
  {author} {\bibfnamefont {I.~D.}\ \bibnamefont {Rodriguez}}, \ and\ \bibinfo
  {author} {\bibfnamefont {J.~I.}\ \bibnamefont {Cirac}},\ }\href {\doibase
  10.1103/PhysRevX.8.011006} {\bibfield  {journal} {\bibinfo  {journal} {Phys.
  Rev. X}\ }\textbf {\bibinfo {volume} {8}},\ \bibinfo {pages} {011006}
  (\bibinfo {year} {2018})}\BibitemShut {NoStop}%
\bibitem [{\citenamefont {Choo}\ \emph {et~al.}(2018)\citenamefont {Choo},
  \citenamefont {Carleo}, \citenamefont {Regnault},\ and\ \citenamefont
  {Neupert}}]{choo_2018}%
  \BibitemOpen
  \bibfield  {author} {\bibinfo {author} {\bibfnamefont {K.}~\bibnamefont
  {Choo}}, \bibinfo {author} {\bibfnamefont {G.}~\bibnamefont {Carleo}},
  \bibinfo {author} {\bibfnamefont {N.}~\bibnamefont {Regnault}}, \ and\
  \bibinfo {author} {\bibfnamefont {T.}~\bibnamefont {Neupert}},\ }\href
  {\doibase 10.1103/PhysRevLett.121.167204} {\bibfield  {journal} {\bibinfo
  {journal} {Phys. Rev. Lett.}\ }\textbf {\bibinfo {volume} {121}},\ \bibinfo
  {pages} {167204} (\bibinfo {year} {2018})}\BibitemShut {NoStop}%
\bibitem [{\citenamefont {Kaubruegger}\ \emph {et~al.}(2018)\citenamefont
  {Kaubruegger}, \citenamefont {Pastori},\ and\ \citenamefont
  {Budich}}]{kaubruegger_2018}%
  \BibitemOpen
  \bibfield  {author} {\bibinfo {author} {\bibfnamefont {R.}~\bibnamefont
  {Kaubruegger}}, \bibinfo {author} {\bibfnamefont {L.}~\bibnamefont
  {Pastori}}, \ and\ \bibinfo {author} {\bibfnamefont {J.~C.}\ \bibnamefont
  {Budich}},\ }\href {\doibase 10.1103/PhysRevB.97.195136} {\bibfield
  {journal} {\bibinfo  {journal} {Phys. Rev. B}\ }\textbf {\bibinfo {volume}
  {97}},\ \bibinfo {pages} {195136} (\bibinfo {year} {2018})}\BibitemShut
  {NoStop}%
\bibitem [{\citenamefont {Levine}\ \emph {et~al.}(2019)\citenamefont {Levine},
  \citenamefont {Sharir}, \citenamefont {Cohen},\ and\ \citenamefont
  {Shashua}}]{levine_2019}%
  \BibitemOpen
  \bibfield  {author} {\bibinfo {author} {\bibfnamefont {Y.}~\bibnamefont
  {Levine}}, \bibinfo {author} {\bibfnamefont {O.}~\bibnamefont {Sharir}},
  \bibinfo {author} {\bibfnamefont {N.}~\bibnamefont {Cohen}}, \ and\ \bibinfo
  {author} {\bibfnamefont {A.}~\bibnamefont {Shashua}},\ }\href {\doibase
  10.1103/PhysRevLett.122.065301} {\bibfield  {journal} {\bibinfo  {journal}
  {Phys. Rev. Lett.}\ }\textbf {\bibinfo {volume} {122}},\ \bibinfo {pages}
  {065301} (\bibinfo {year} {2019})}\BibitemShut {NoStop}%
\bibitem [{\citenamefont {Sharir}\ \emph {et~al.}(2019)\citenamefont {Sharir},
  \citenamefont {Levine}, \citenamefont {Wies}, \citenamefont {Carleo},\ and\
  \citenamefont {Shashua}}]{sharir_2019}%
  \BibitemOpen
  \bibfield  {author} {\bibinfo {author} {\bibfnamefont {O.}~\bibnamefont
  {Sharir}}, \bibinfo {author} {\bibfnamefont {Y.}~\bibnamefont {Levine}},
  \bibinfo {author} {\bibfnamefont {N.}~\bibnamefont {Wies}}, \bibinfo {author}
  {\bibfnamefont {G.}~\bibnamefont {Carleo}}, \ and\ \bibinfo {author}
  {\bibfnamefont {A.}~\bibnamefont {Shashua}},\ }\href@noop {} {\bibfield
  {journal} {\bibinfo  {journal} {arXiv:1902.04057}\ } (\bibinfo {year}
  {2019})}\BibitemShut {NoStop}%
\bibitem [{\citenamefont {Barreiro}\ \emph {et~al.}(2011)\citenamefont
  {Barreiro}, \citenamefont {M{\"u}ller}, \citenamefont {Schindler},
  \citenamefont {Nigg}, \citenamefont {Monz}, \citenamefont {Chwalla},
  \citenamefont {Hennrich}, \citenamefont {Roos}, \citenamefont {Zoller},\ and\
  \citenamefont {Blatt}}]{barreiro_2011}%
  \BibitemOpen
  \bibfield  {author} {\bibinfo {author} {\bibfnamefont {J.~T.}\ \bibnamefont
  {Barreiro}}, \bibinfo {author} {\bibfnamefont {M.}~\bibnamefont
  {M{\"u}ller}}, \bibinfo {author} {\bibfnamefont {P.}~\bibnamefont
  {Schindler}}, \bibinfo {author} {\bibfnamefont {D.}~\bibnamefont {Nigg}},
  \bibinfo {author} {\bibfnamefont {T.}~\bibnamefont {Monz}}, \bibinfo {author}
  {\bibfnamefont {M.}~\bibnamefont {Chwalla}}, \bibinfo {author} {\bibfnamefont
  {M.}~\bibnamefont {Hennrich}}, \bibinfo {author} {\bibfnamefont {C.~F.}\
  \bibnamefont {Roos}}, \bibinfo {author} {\bibfnamefont {P.}~\bibnamefont
  {Zoller}}, \ and\ \bibinfo {author} {\bibfnamefont {R.}~\bibnamefont
  {Blatt}},\ }\href@noop {} {\bibfield  {journal} {\bibinfo  {journal}
  {Nature}\ }\textbf {\bibinfo {volume} {470}},\ \bibinfo {pages} {486}
  (\bibinfo {year} {2011})}\BibitemShut {NoStop}%
\bibitem [{\citenamefont {Barontini}\ \emph {et~al.}(2013)\citenamefont
  {Barontini}, \citenamefont {Labouvie}, \citenamefont {Stubenrauch},
  \citenamefont {Vogler}, \citenamefont {Guarrera},\ and\ \citenamefont
  {Ott}}]{barontini_2013}%
  \BibitemOpen
  \bibfield  {author} {\bibinfo {author} {\bibfnamefont {G.}~\bibnamefont
  {Barontini}}, \bibinfo {author} {\bibfnamefont {R.}~\bibnamefont {Labouvie}},
  \bibinfo {author} {\bibfnamefont {F.}~\bibnamefont {Stubenrauch}}, \bibinfo
  {author} {\bibfnamefont {A.}~\bibnamefont {Vogler}}, \bibinfo {author}
  {\bibfnamefont {V.}~\bibnamefont {Guarrera}}, \ and\ \bibinfo {author}
  {\bibfnamefont {H.}~\bibnamefont {Ott}},\ }\href {\doibase
  10.1103/PhysRevLett.110.035302} {\bibfield  {journal} {\bibinfo  {journal}
  {Phys. Rev. Lett.}\ }\textbf {\bibinfo {volume} {110}},\ \bibinfo {pages}
  {035302} (\bibinfo {year} {2013})}\BibitemShut {NoStop}%
\bibitem [{\citenamefont {Labouvie}\ \emph {et~al.}(2016)\citenamefont
  {Labouvie}, \citenamefont {Santra}, \citenamefont {Heun},\ and\ \citenamefont
  {Ott}}]{labouvie_2016}%
  \BibitemOpen
  \bibfield  {author} {\bibinfo {author} {\bibfnamefont {R.}~\bibnamefont
  {Labouvie}}, \bibinfo {author} {\bibfnamefont {B.}~\bibnamefont {Santra}},
  \bibinfo {author} {\bibfnamefont {S.}~\bibnamefont {Heun}}, \ and\ \bibinfo
  {author} {\bibfnamefont {H.}~\bibnamefont {Ott}},\ }\href {\doibase
  10.1103/PhysRevLett.116.235302} {\bibfield  {journal} {\bibinfo  {journal}
  {Phys. Rev. Lett.}\ }\textbf {\bibinfo {volume} {116}},\ \bibinfo {pages}
  {235302} (\bibinfo {year} {2016})}\BibitemShut {NoStop}%
\bibitem [{\citenamefont {Tomita}\ \emph {et~al.}(2017)\citenamefont {Tomita},
  \citenamefont {Nakajima}, \citenamefont {Danshita}, \citenamefont {Takasu},\
  and\ \citenamefont {Takahashi}}]{tomita_2017}%
  \BibitemOpen
  \bibfield  {author} {\bibinfo {author} {\bibfnamefont {T.}~\bibnamefont
  {Tomita}}, \bibinfo {author} {\bibfnamefont {S.}~\bibnamefont {Nakajima}},
  \bibinfo {author} {\bibfnamefont {I.}~\bibnamefont {Danshita}}, \bibinfo
  {author} {\bibfnamefont {Y.}~\bibnamefont {Takasu}}, \ and\ \bibinfo {author}
  {\bibfnamefont {Y.}~\bibnamefont {Takahashi}},\ }\href@noop {} {\bibfield
  {journal} {\bibinfo  {journal} {Science Advances}\ }\textbf {\bibinfo
  {volume} {3}},\ \bibinfo {pages} {e1701513} (\bibinfo {year}
  {2017})}\BibitemShut {NoStop}%
\bibitem [{\citenamefont {Fitzpatrick}\ \emph {et~al.}(2017)\citenamefont
  {Fitzpatrick}, \citenamefont {Sundaresan}, \citenamefont {Li}, \citenamefont
  {Koch},\ and\ \citenamefont {Houck}}]{fitzpatrick_2017}%
  \BibitemOpen
  \bibfield  {author} {\bibinfo {author} {\bibfnamefont {M.}~\bibnamefont
  {Fitzpatrick}}, \bibinfo {author} {\bibfnamefont {N.~M.}\ \bibnamefont
  {Sundaresan}}, \bibinfo {author} {\bibfnamefont {A.~C.~Y.}\ \bibnamefont
  {Li}}, \bibinfo {author} {\bibfnamefont {J.}~\bibnamefont {Koch}}, \ and\
  \bibinfo {author} {\bibfnamefont {A.~A.}\ \bibnamefont {Houck}},\ }\href
  {\doibase 10.1103/PhysRevX.7.011016} {\bibfield  {journal} {\bibinfo
  {journal} {Phys. Rev. X}\ }\textbf {\bibinfo {volume} {7}},\ \bibinfo {pages}
  {011016} (\bibinfo {year} {2017})}\BibitemShut {NoStop}%
\bibitem [{\citenamefont {Lindblad}(1976)}]{lindblad_1976}%
  \BibitemOpen
  \bibfield  {author} {\bibinfo {author} {\bibfnamefont {G.}~\bibnamefont
  {Lindblad}},\ }\href {https://projecteuclid.org:443/euclid.cmp/1103899849}
  {\bibfield  {journal} {\bibinfo  {journal} {Comm. Math. Phys.}\ }\textbf
  {\bibinfo {volume} {48}},\ \bibinfo {pages} {119} (\bibinfo {year}
  {1976})}\BibitemShut {NoStop}%
\bibitem [{\citenamefont {Kraus}\ \emph {et~al.}(2008)\citenamefont {Kraus},
  \citenamefont {B\"uchler}, \citenamefont {Diehl}, \citenamefont {Kantian},
  \citenamefont {Micheli},\ and\ \citenamefont {Zoller}}]{kraus_2008}%
  \BibitemOpen
  \bibfield  {author} {\bibinfo {author} {\bibfnamefont {B.}~\bibnamefont
  {Kraus}}, \bibinfo {author} {\bibfnamefont {H.~P.}\ \bibnamefont
  {B\"uchler}}, \bibinfo {author} {\bibfnamefont {S.}~\bibnamefont {Diehl}},
  \bibinfo {author} {\bibfnamefont {A.}~\bibnamefont {Kantian}}, \bibinfo
  {author} {\bibfnamefont {A.}~\bibnamefont {Micheli}}, \ and\ \bibinfo
  {author} {\bibfnamefont {P.}~\bibnamefont {Zoller}},\ }\href {\doibase
  10.1103/PhysRevA.78.042307} {\bibfield  {journal} {\bibinfo  {journal} {Phys.
  Rev. A}\ }\textbf {\bibinfo {volume} {78}},\ \bibinfo {pages} {042307}
  (\bibinfo {year} {2008})}\BibitemShut {NoStop}%
\bibitem [{\citenamefont {Kastoryano}\ \emph {et~al.}(2011)\citenamefont
  {Kastoryano}, \citenamefont {Reiter},\ and\ \citenamefont
  {S\o{}rensen}}]{kastoryano_2011}%
  \BibitemOpen
  \bibfield  {author} {\bibinfo {author} {\bibfnamefont {M.~J.}\ \bibnamefont
  {Kastoryano}}, \bibinfo {author} {\bibfnamefont {F.}~\bibnamefont {Reiter}},
  \ and\ \bibinfo {author} {\bibfnamefont {A.~S.}\ \bibnamefont
  {S\o{}rensen}},\ }\href {\doibase 10.1103/PhysRevLett.106.090502} {\bibfield
  {journal} {\bibinfo  {journal} {Phys. Rev. Lett.}\ }\textbf {\bibinfo
  {volume} {106}},\ \bibinfo {pages} {090502} (\bibinfo {year}
  {2011})}\BibitemShut {NoStop}%
\bibitem [{\citenamefont {Diehl}\ \emph {et~al.}(2011)\citenamefont {Diehl},
  \citenamefont {Rico}, \citenamefont {Baranov},\ and\ \citenamefont
  {Zoller}}]{diehl_2011}%
  \BibitemOpen
  \bibfield  {author} {\bibinfo {author} {\bibfnamefont {S.}~\bibnamefont
  {Diehl}}, \bibinfo {author} {\bibfnamefont {E.}~\bibnamefont {Rico}},
  \bibinfo {author} {\bibfnamefont {M.~A.}\ \bibnamefont {Baranov}}, \ and\
  \bibinfo {author} {\bibfnamefont {P.}~\bibnamefont {Zoller}},\ }\href@noop {}
  {\bibfield  {journal} {\bibinfo  {journal} {Nature Physics}\ }\textbf
  {\bibinfo {volume} {7}},\ \bibinfo {pages} {971} (\bibinfo {year}
  {2011})}\BibitemShut {NoStop}%
\bibitem [{\citenamefont {Bardyn}\ \emph {et~al.}(2013)\citenamefont {Bardyn},
  \citenamefont {Baranov}, \citenamefont {Kraus}, \citenamefont {Rico},
  \citenamefont {{\.I}mamo{\u{g}}lu}, \citenamefont {Zoller},\ and\
  \citenamefont {Diehl}}]{bardyn_2013}%
  \BibitemOpen
  \bibfield  {author} {\bibinfo {author} {\bibfnamefont {C.}~\bibnamefont
  {Bardyn}}, \bibinfo {author} {\bibfnamefont {M.}~\bibnamefont {Baranov}},
  \bibinfo {author} {\bibfnamefont {C.}~\bibnamefont {Kraus}}, \bibinfo
  {author} {\bibfnamefont {E.}~\bibnamefont {Rico}}, \bibinfo {author}
  {\bibfnamefont {A.}~\bibnamefont {{\.I}mamo{\u{g}}lu}}, \bibinfo {author}
  {\bibfnamefont {P.}~\bibnamefont {Zoller}}, \ and\ \bibinfo {author}
  {\bibfnamefont {S.}~\bibnamefont {Diehl}},\ }\href@noop {} {\bibfield
  {journal} {\bibinfo  {journal} {New Journal of Physics}\ }\textbf {\bibinfo
  {volume} {15}},\ \bibinfo {pages} {085001} (\bibinfo {year}
  {2013})}\BibitemShut {NoStop}%
\bibitem [{\citenamefont {Diehl}\ \emph {et~al.}(2010)\citenamefont {Diehl},
  \citenamefont {Tomadin}, \citenamefont {Micheli}, \citenamefont {Fazio},\
  and\ \citenamefont {Zoller}}]{diehl_2010}%
  \BibitemOpen
  \bibfield  {author} {\bibinfo {author} {\bibfnamefont {S.}~\bibnamefont
  {Diehl}}, \bibinfo {author} {\bibfnamefont {A.}~\bibnamefont {Tomadin}},
  \bibinfo {author} {\bibfnamefont {A.}~\bibnamefont {Micheli}}, \bibinfo
  {author} {\bibfnamefont {R.}~\bibnamefont {Fazio}}, \ and\ \bibinfo {author}
  {\bibfnamefont {P.}~\bibnamefont {Zoller}},\ }\href {\doibase
  10.1103/PhysRevLett.105.015702} {\bibfield  {journal} {\bibinfo  {journal}
  {Phys. Rev. Lett.}\ }\textbf {\bibinfo {volume} {105}},\ \bibinfo {pages}
  {015702} (\bibinfo {year} {2010})}\BibitemShut {NoStop}%
\bibitem [{\citenamefont {Tomadin}\ \emph {et~al.}(2011)\citenamefont
  {Tomadin}, \citenamefont {Diehl},\ and\ \citenamefont
  {Zoller}}]{tomadin_2011}%
  \BibitemOpen
  \bibfield  {author} {\bibinfo {author} {\bibfnamefont {A.}~\bibnamefont
  {Tomadin}}, \bibinfo {author} {\bibfnamefont {S.}~\bibnamefont {Diehl}}, \
  and\ \bibinfo {author} {\bibfnamefont {P.}~\bibnamefont {Zoller}},\ }\href
  {\doibase 10.1103/PhysRevA.83.013611} {\bibfield  {journal} {\bibinfo
  {journal} {Phys. Rev. A}\ }\textbf {\bibinfo {volume} {83}},\ \bibinfo
  {pages} {013611} (\bibinfo {year} {2011})}\BibitemShut {NoStop}%
\bibitem [{\citenamefont {Ates}\ \emph {et~al.}(2012)\citenamefont {Ates},
  \citenamefont {Olmos}, \citenamefont {Garrahan},\ and\ \citenamefont
  {Lesanovsky}}]{ates_2012}%
  \BibitemOpen
  \bibfield  {author} {\bibinfo {author} {\bibfnamefont {C.}~\bibnamefont
  {Ates}}, \bibinfo {author} {\bibfnamefont {B.}~\bibnamefont {Olmos}},
  \bibinfo {author} {\bibfnamefont {J.~P.}\ \bibnamefont {Garrahan}}, \ and\
  \bibinfo {author} {\bibfnamefont {I.}~\bibnamefont {Lesanovsky}},\ }\href
  {\doibase 10.1103/PhysRevA.85.043620} {\bibfield  {journal} {\bibinfo
  {journal} {Phys. Rev. A}\ }\textbf {\bibinfo {volume} {85}},\ \bibinfo
  {pages} {043620} (\bibinfo {year} {2012})}\BibitemShut {NoStop}%
\bibitem [{\citenamefont {Gong}\ \emph {et~al.}(2018)\citenamefont {Gong},
  \citenamefont {Hamazaki},\ and\ \citenamefont {Ueda}}]{gong_2018}%
  \BibitemOpen
  \bibfield  {author} {\bibinfo {author} {\bibfnamefont {Z.}~\bibnamefont
  {Gong}}, \bibinfo {author} {\bibfnamefont {R.}~\bibnamefont {Hamazaki}}, \
  and\ \bibinfo {author} {\bibfnamefont {M.}~\bibnamefont {Ueda}},\ }\href
  {\doibase 10.1103/PhysRevLett.120.040404} {\bibfield  {journal} {\bibinfo
  {journal} {Phys. Rev. Lett.}\ }\textbf {\bibinfo {volume} {120}},\ \bibinfo
  {pages} {040404} (\bibinfo {year} {2018})}\BibitemShut {NoStop}%
\bibitem [{\citenamefont {Gambetta}\ \emph {et~al.}(2019)\citenamefont
  {Gambetta}, \citenamefont {Carollo}, \citenamefont {Marcuzzi}, \citenamefont
  {Garrahan},\ and\ \citenamefont {Lesanovsky}}]{gambetta_2019}%
  \BibitemOpen
  \bibfield  {author} {\bibinfo {author} {\bibfnamefont {F.~M.}\ \bibnamefont
  {Gambetta}}, \bibinfo {author} {\bibfnamefont {F.}~\bibnamefont {Carollo}},
  \bibinfo {author} {\bibfnamefont {M.}~\bibnamefont {Marcuzzi}}, \bibinfo
  {author} {\bibfnamefont {J.~P.}\ \bibnamefont {Garrahan}}, \ and\ \bibinfo
  {author} {\bibfnamefont {I.}~\bibnamefont {Lesanovsky}},\ }\href {\doibase
  10.1103/PhysRevLett.122.015701} {\bibfield  {journal} {\bibinfo  {journal}
  {Phys. Rev. Lett.}\ }\textbf {\bibinfo {volume} {122}},\ \bibinfo {pages}
  {015701} (\bibinfo {year} {2019})}\BibitemShut {NoStop}%
\bibitem [{\citenamefont {Cui}\ \emph {et~al.}(2015)\citenamefont {Cui},
  \citenamefont {Cirac},\ and\ \citenamefont {Ba{\~{n}}uls}}]{cui_2015}%
  \BibitemOpen
  \bibfield  {author} {\bibinfo {author} {\bibfnamefont {J.}~\bibnamefont
  {Cui}}, \bibinfo {author} {\bibfnamefont {J.~I.}\ \bibnamefont {Cirac}}, \
  and\ \bibinfo {author} {\bibfnamefont {M.~C.}\ \bibnamefont {Ba{\~{n}}uls}},\
  }\href {\doibase 10.1103/PhysRevLett.114.220601} {\bibfield  {journal}
  {\bibinfo  {journal} {Phys. Rev. Lett.}\ }\textbf {\bibinfo {volume} {114}},\
  \bibinfo {pages} {220601} (\bibinfo {year} {2015})}\BibitemShut {NoStop}%
\bibitem [{\citenamefont {Prosen}\ and\ \citenamefont
  {Pi\v{z}orn}(2007)}]{prosen_2007}%
  \BibitemOpen
  \bibfield  {author} {\bibinfo {author} {\bibfnamefont {T.}~\bibnamefont
  {Prosen}}\ and\ \bibinfo {author} {\bibfnamefont {I.}~\bibnamefont
  {Pi\v{z}orn}},\ }\href {\doibase 10.1103/PhysRevA.76.032316} {\bibfield
  {journal} {\bibinfo  {journal} {Phys. Rev. A}\ }\textbf {\bibinfo {volume}
  {76}},\ \bibinfo {pages} {032316} (\bibinfo {year} {2007})}\BibitemShut
  {NoStop}%
\bibitem [{\citenamefont {Pi\v{z}orn}\ and\ \citenamefont
  {Prosen}(2009)}]{pizorn_2009}%
  \BibitemOpen
  \bibfield  {author} {\bibinfo {author} {\bibfnamefont {I.}~\bibnamefont
  {Pi\v{z}orn}}\ and\ \bibinfo {author} {\bibfnamefont {T.}~\bibnamefont
  {Prosen}},\ }\href {\doibase 10.1103/PhysRevB.79.184416} {\bibfield
  {journal} {\bibinfo  {journal} {Phys. Rev. B}\ }\textbf {\bibinfo {volume}
  {79}},\ \bibinfo {pages} {184416} (\bibinfo {year} {2009})}\BibitemShut
  {NoStop}%
\bibitem [{\citenamefont {Cai}\ and\ \citenamefont {Barthel}(2013)}]{cai_2013}%
  \BibitemOpen
  \bibfield  {author} {\bibinfo {author} {\bibfnamefont {Z.}~\bibnamefont
  {Cai}}\ and\ \bibinfo {author} {\bibfnamefont {T.}~\bibnamefont {Barthel}},\
  }\href {\doibase 10.1103/PhysRevLett.111.150403} {\bibfield  {journal}
  {\bibinfo  {journal} {Phys. Rev. Lett.}\ }\textbf {\bibinfo {volume} {111}},\
  \bibinfo {pages} {150403} (\bibinfo {year} {2013})}\BibitemShut {NoStop}%
\bibitem [{\citenamefont {Werner}\ \emph {et~al.}(2016)\citenamefont {Werner},
  \citenamefont {Jaschke}, \citenamefont {Silvi}, \citenamefont {Kliesch},
  \citenamefont {Calarco}, \citenamefont {Eisert},\ and\ \citenamefont
  {Montangero}}]{werner_2016}%
  \BibitemOpen
  \bibfield  {author} {\bibinfo {author} {\bibfnamefont {A.~H.}\ \bibnamefont
  {Werner}}, \bibinfo {author} {\bibfnamefont {D.}~\bibnamefont {Jaschke}},
  \bibinfo {author} {\bibfnamefont {P.}~\bibnamefont {Silvi}}, \bibinfo
  {author} {\bibfnamefont {M.}~\bibnamefont {Kliesch}}, \bibinfo {author}
  {\bibfnamefont {T.}~\bibnamefont {Calarco}}, \bibinfo {author} {\bibfnamefont
  {J.}~\bibnamefont {Eisert}}, \ and\ \bibinfo {author} {\bibfnamefont
  {S.}~\bibnamefont {Montangero}},\ }\href {\doibase
  10.1103/PhysRevLett.116.237201} {\bibfield  {journal} {\bibinfo  {journal}
  {Phys. Rev. Lett.}\ }\textbf {\bibinfo {volume} {116}},\ \bibinfo {pages}
  {237201} (\bibinfo {year} {2016})}\BibitemShut {NoStop}%
\bibitem [{\citenamefont {Gangat}\ \emph {et~al.}(2017)\citenamefont {Gangat},
  \citenamefont {I},\ and\ \citenamefont {Kao}}]{gangat_2017}%
  \BibitemOpen
  \bibfield  {author} {\bibinfo {author} {\bibfnamefont {A.~A.}\ \bibnamefont
  {Gangat}}, \bibinfo {author} {\bibfnamefont {T.}~\bibnamefont {I}}, \ and\
  \bibinfo {author} {\bibfnamefont {Y.-J.}\ \bibnamefont {Kao}},\ }\href
  {\doibase 10.1103/PhysRevLett.119.010501} {\bibfield  {journal} {\bibinfo
  {journal} {Phys. Rev. Lett.}\ }\textbf {\bibinfo {volume} {119}},\ \bibinfo
  {pages} {010501} (\bibinfo {year} {2017})}\BibitemShut {NoStop}%
\bibitem [{\citenamefont {Kshetrimayum}\ \emph {et~al.}(2017)\citenamefont
  {Kshetrimayum}, \citenamefont {Weimer},\ and\ \citenamefont
  {Or{\'u}s}}]{kshetrimayum_2017}%
  \BibitemOpen
  \bibfield  {author} {\bibinfo {author} {\bibfnamefont {A.}~\bibnamefont
  {Kshetrimayum}}, \bibinfo {author} {\bibfnamefont {H.}~\bibnamefont
  {Weimer}}, \ and\ \bibinfo {author} {\bibfnamefont {R.}~\bibnamefont
  {Or{\'u}s}},\ }\href@noop {} {\bibfield  {journal} {\bibinfo  {journal}
  {Nature communications}\ }\textbf {\bibinfo {volume} {8}},\ \bibinfo {pages}
  {1291} (\bibinfo {year} {2017})}\BibitemShut {NoStop}%
\bibitem [{\citenamefont {Weimer}(2015)}]{weimer_2015}%
  \BibitemOpen
  \bibfield  {author} {\bibinfo {author} {\bibfnamefont {H.}~\bibnamefont
  {Weimer}},\ }\href {\doibase 10.1103/PhysRevLett.114.040402} {\bibfield
  {journal} {\bibinfo  {journal} {Phys. Rev. Lett.}\ }\textbf {\bibinfo
  {volume} {114}},\ \bibinfo {pages} {040402} (\bibinfo {year}
  {2015})}\BibitemShut {NoStop}%
\bibitem [{\citenamefont {Jin}\ \emph {et~al.}(2016)\citenamefont {Jin},
  \citenamefont {Biella}, \citenamefont {Viyuela}, \citenamefont {Mazza},
  \citenamefont {Keeling}, \citenamefont {Fazio},\ and\ \citenamefont
  {Rossini}}]{jiasen_2016}%
  \BibitemOpen
  \bibfield  {author} {\bibinfo {author} {\bibfnamefont {J.}~\bibnamefont
  {Jin}}, \bibinfo {author} {\bibfnamefont {A.}~\bibnamefont {Biella}},
  \bibinfo {author} {\bibfnamefont {O.}~\bibnamefont {Viyuela}}, \bibinfo
  {author} {\bibfnamefont {L.}~\bibnamefont {Mazza}}, \bibinfo {author}
  {\bibfnamefont {J.}~\bibnamefont {Keeling}}, \bibinfo {author} {\bibfnamefont
  {R.}~\bibnamefont {Fazio}}, \ and\ \bibinfo {author} {\bibfnamefont
  {D.}~\bibnamefont {Rossini}},\ }\href {\doibase 10.1103/PhysRevX.6.031011}
  {\bibfield  {journal} {\bibinfo  {journal} {Phys. Rev. X}\ }\textbf {\bibinfo
  {volume} {6}},\ \bibinfo {pages} {031011} (\bibinfo {year}
  {2016})}\BibitemShut {NoStop}%
\bibitem [{\citenamefont {Jin}\ \emph {et~al.}(2018)\citenamefont {Jin},
  \citenamefont {Biella}, \citenamefont {Viyuela}, \citenamefont {Ciuti},
  \citenamefont {Fazio},\ and\ \citenamefont {Rossini}}]{jin_2018}%
  \BibitemOpen
  \bibfield  {author} {\bibinfo {author} {\bibfnamefont {J.}~\bibnamefont
  {Jin}}, \bibinfo {author} {\bibfnamefont {A.}~\bibnamefont {Biella}},
  \bibinfo {author} {\bibfnamefont {O.}~\bibnamefont {Viyuela}}, \bibinfo
  {author} {\bibfnamefont {C.}~\bibnamefont {Ciuti}}, \bibinfo {author}
  {\bibfnamefont {R.}~\bibnamefont {Fazio}}, \ and\ \bibinfo {author}
  {\bibfnamefont {D.}~\bibnamefont {Rossini}},\ }\href {\doibase
  10.1103/PhysRevB.98.241108} {\bibfield  {journal} {\bibinfo  {journal} {Phys.
  Rev. B}\ }\textbf {\bibinfo {volume} {98}},\ \bibinfo {pages} {241108}
  (\bibinfo {year} {2018})}\BibitemShut {NoStop}%
\bibitem [{\citenamefont {Rivas}\ and\ \citenamefont
  {Huelga}(2011)}]{rivas_2011}%
  \BibitemOpen
  \bibfield  {author} {\bibinfo {author} {\bibfnamefont {{\'A}.}~\bibnamefont
  {Rivas}}\ and\ \bibinfo {author} {\bibfnamefont {S.}~\bibnamefont {Huelga}},\
  }\href {https://books.google.co.jp/books?id=FGCuYsIZAA0C} {\emph {\bibinfo
  {title} {Open Quantum Systems: An Introduction}}},\ SpringerBriefs in
  Physics\ (\bibinfo  {publisher} {Springer Berlin Heidelberg},\ \bibinfo
  {year} {2011})\BibitemShut {NoStop}%
\bibitem [{\citenamefont {Breuer}\ \emph {et~al.}(2002)\citenamefont {Breuer},
  \citenamefont {Petruccione} \emph {et~al.}}]{breuer_2002}%
  \BibitemOpen
  \bibfield  {author} {\bibinfo {author} {\bibfnamefont {H.-P.}\ \bibnamefont
  {Breuer}}, \bibinfo {author} {\bibfnamefont {F.}~\bibnamefont {Petruccione}},
   \emph {et~al.},\ }\href@noop {} {\emph {\bibinfo {title} {The theory of open
  quantum systems}}}\ (\bibinfo  {publisher} {Oxford University Press on
  Demand},\ \bibinfo {year} {2002})\BibitemShut {NoStop}%
\bibitem [{\citenamefont {Schirmer}\ and\ \citenamefont
  {Wang}(2010)}]{schirmer_2010}%
  \BibitemOpen
  \bibfield  {author} {\bibinfo {author} {\bibfnamefont {S.~G.}\ \bibnamefont
  {Schirmer}}\ and\ \bibinfo {author} {\bibfnamefont {X.}~\bibnamefont
  {Wang}},\ }\href {\doibase 10.1103/PhysRevA.81.062306} {\bibfield  {journal}
  {\bibinfo  {journal} {Phys. Rev. A}\ }\textbf {\bibinfo {volume} {81}},\
  \bibinfo {pages} {062306} (\bibinfo {year} {2010})}\BibitemShut {NoStop}%
\bibitem [{\citenamefont {Prosen}(2012)}]{prosen_2012}%
  \BibitemOpen
  \bibfield  {author} {\bibinfo {author} {\bibfnamefont {T.}~\bibnamefont
  {Prosen}},\ }\href@noop {} {\bibfield  {journal} {\bibinfo  {journal}
  {Physica Scripta}\ }\textbf {\bibinfo {volume} {86}},\ \bibinfo {pages}
  {058511} (\bibinfo {year} {2012})}\BibitemShut {NoStop}%
\bibitem [{\citenamefont {Horstmann}\ \emph {et~al.}(2013)\citenamefont
  {Horstmann}, \citenamefont {Cirac},\ and\ \citenamefont
  {Giedke}}]{horstmann_2013}%
  \BibitemOpen
  \bibfield  {author} {\bibinfo {author} {\bibfnamefont {B.}~\bibnamefont
  {Horstmann}}, \bibinfo {author} {\bibfnamefont {J.~I.}\ \bibnamefont
  {Cirac}}, \ and\ \bibinfo {author} {\bibfnamefont {G.}~\bibnamefont
  {Giedke}},\ }\href {\doibase 10.1103/PhysRevA.87.012108} {\bibfield
  {journal} {\bibinfo  {journal} {Phys. Rev. A}\ }\textbf {\bibinfo {volume}
  {87}},\ \bibinfo {pages} {012108} (\bibinfo {year} {2013})}\BibitemShut
  {NoStop}%
\bibitem [{\citenamefont {Sorella}(2001)}]{sorella_2001}%
  \BibitemOpen
  \bibfield  {author} {\bibinfo {author} {\bibfnamefont {S.}~\bibnamefont
  {Sorella}},\ }\href {\doibase 10.1103/PhysRevB.64.024512} {\bibfield
  {journal} {\bibinfo  {journal} {Phys. Rev. B}\ }\textbf {\bibinfo {volume}
  {64}},\ \bibinfo {pages} {024512} (\bibinfo {year} {2001})}\BibitemShut
  {NoStop}%
\bibitem [{\citenamefont {Amari}\ \emph {et~al.}(1992)\citenamefont {Amari},
  \citenamefont {Kurata},\ and\ \citenamefont {H}}]{amari_1992}%
  \BibitemOpen
  \bibfield  {author} {\bibinfo {author} {\bibfnamefont {S.-I.}\ \bibnamefont
  {Amari}}, \bibinfo {author} {\bibfnamefont {K.}~\bibnamefont {Kurata}}, \
  and\ \bibinfo {author} {\bibfnamefont {N.}~\bibnamefont {H}},\ }\href@noop {}
  {\bibfield  {journal} {\bibinfo  {journal} {IEEE Transactions on Neural
  Networks}\ }\textbf {\bibinfo {volume} {3}},\ \bibinfo {pages} {260}
  (\bibinfo {year} {1992})}\BibitemShut {NoStop}%
\bibitem [{\citenamefont {Amari}(1998)}]{amari_1998}%
  \BibitemOpen
  \bibfield  {author} {\bibinfo {author} {\bibfnamefont {S.-I.}\ \bibnamefont
  {Amari}},\ }\href {\doibase 10.1162/089976698300017746} {\bibfield  {journal}
  {\bibinfo  {journal} {Neural Computation}\ }\textbf {\bibinfo {volume}
  {10}},\ \bibinfo {pages} {251} (\bibinfo {year} {1998})}\BibitemShut
  {NoStop}%
\bibitem [{\citenamefont {Lee}\ \emph {et~al.}(2013)\citenamefont {Lee},
  \citenamefont {Gopalakrishnan},\ and\ \citenamefont {Lukin}}]{lee_2013}%
  \BibitemOpen
  \bibfield  {author} {\bibinfo {author} {\bibfnamefont {T.~E.}\ \bibnamefont
  {Lee}}, \bibinfo {author} {\bibfnamefont {S.}~\bibnamefont {Gopalakrishnan}},
  \ and\ \bibinfo {author} {\bibfnamefont {M.~D.}\ \bibnamefont {Lukin}},\
  }\href {\doibase 10.1103/PhysRevLett.110.257204} {\bibfield  {journal}
  {\bibinfo  {journal} {Phys. Rev. Lett.}\ }\textbf {\bibinfo {volume} {110}},\
  \bibinfo {pages} {257204} (\bibinfo {year} {2013})}\BibitemShut {NoStop}%
\bibitem [{\citenamefont {Torlai}\ and\ \citenamefont
  {Melko}(2018)}]{torlai_2018}%
  \BibitemOpen
  \bibfield  {author} {\bibinfo {author} {\bibfnamefont {G.}~\bibnamefont
  {Torlai}}\ and\ \bibinfo {author} {\bibfnamefont {R.~G.}\ \bibnamefont
  {Melko}},\ }\href {\doibase 10.1103/PhysRevLett.120.240503} {\bibfield
  {journal} {\bibinfo  {journal} {Phys. Rev. Lett.}\ }\textbf {\bibinfo
  {volume} {120}},\ \bibinfo {pages} {240503} (\bibinfo {year}
  {2018})}\BibitemShut {NoStop}%
\bibitem [{\citenamefont {Lehoucq}\ \emph {et~al.}(1998)\citenamefont
  {Lehoucq}, \citenamefont {Sorensen},\ and\ \citenamefont
  {Yang}}]{arpack_1998}%
  \BibitemOpen
  \bibfield  {author} {\bibinfo {author} {\bibfnamefont {R.~B.}\ \bibnamefont
  {Lehoucq}}, \bibinfo {author} {\bibfnamefont {D.~C.}\ \bibnamefont
  {Sorensen}}, \ and\ \bibinfo {author} {\bibfnamefont {C.}~\bibnamefont
  {Yang}},\ }\href@noop {} {\bibfield  {journal} {\bibinfo  {journal} {{\it
  ARPACK users' guide: solution of large-scale eigenvalue problems with
  implicitly restarted Arnoldi methods}}\ }\textbf {\bibinfo {volume} {6}}
  (\bibinfo {year} {1998})}\BibitemShut {NoStop}%
\bibitem [{\citenamefont {Jozsa}(1994)}]{jozsa_1994}%
  \BibitemOpen
  \bibfield  {author} {\bibinfo {author} {\bibfnamefont {R.}~\bibnamefont
  {Jozsa}},\ }\href {\doibase 10.1080/09500349414552171} {\bibfield  {journal}
  {\bibinfo  {journal} {Journal of Modern Optics}\ }\textbf {\bibinfo {volume}
  {41}},\ \bibinfo {pages} {2315} (\bibinfo {year} {1994})}\BibitemShut
  {NoStop}%
\bibitem [{Note1()}]{Note1}%
  \BibitemOpen
  \bibinfo {note} {We can show that $|\mathinner {\delimiter "426830A
  {\mathinner {\delimiter "426830A {{\rho }_\protect \mathrm {SS}|{\rho
  }_\protect \mathrm {RBM}}\delimiter "526930B }}\delimiter "526930B }|^2\geq
  1-\protect \frac {\delimiter "426830A \mathinner {\delimiter "426830A
  {\protect \mathaccentV {hat}05E{\protect \mathcal {L}}^{\dagger } \protect
  \mathaccentV {hat}05E{\protect \mathcal {L}}}\delimiter "526930B }\delimiter
  "526930B }{\Delta }$, where $\Delta $ is the spectral gap of $\protect
  \mathaccentV {hat}05E{\protect \mathcal {L}}^{\dagger } \protect \mathaccentV
  {hat}05E{\protect \mathcal {L}}$}\BibitemShut {NoStop}%
\bibitem [{\citenamefont {Hartmann}\ and\ \citenamefont
  {Carleo}(2019)}]{carleo_2019}%
  \BibitemOpen
  \bibfield  {author} {\bibinfo {author} {\bibfnamefont {M.~J.}\ \bibnamefont
  {Hartmann}}\ and\ \bibinfo {author} {\bibfnamefont {G.}~\bibnamefont
  {Carleo}},\ }\href@noop {} {\bibfield  {journal} {\bibinfo  {journal}
  {arXiv:1902.05131}\ } (\bibinfo {year} {2019})}\BibitemShut {NoStop}%
\bibitem [{\citenamefont {Nagy}\ and\ \citenamefont
  {Savona}(2019)}]{nagy_2019}%
  \BibitemOpen
  \bibfield  {author} {\bibinfo {author} {\bibfnamefont {A.}~\bibnamefont
  {Nagy}}\ and\ \bibinfo {author} {\bibfnamefont {V.}~\bibnamefont {Savona}},\
  }\href@noop {} {\bibfield  {journal} {\bibinfo  {journal} {arXiv preprint
  arXiv:1902.09483}\ } (\bibinfo {year} {2019})}\BibitemShut {NoStop}%
\bibitem [{\citenamefont {Vicentini}\ \emph {et~al.}(2019)\citenamefont
  {Vicentini}, \citenamefont {Biella}, \citenamefont {Regnault},\ and\
  \citenamefont {Ciuti}}]{vicentini_2019}%
  \BibitemOpen
  \bibfield  {author} {\bibinfo {author} {\bibfnamefont {F.}~\bibnamefont
  {Vicentini}}, \bibinfo {author} {\bibfnamefont {A.}~\bibnamefont {Biella}},
  \bibinfo {author} {\bibfnamefont {N.}~\bibnamefont {Regnault}}, \ and\
  \bibinfo {author} {\bibfnamefont {C.}~\bibnamefont {Ciuti}},\ }\href@noop {}
  {\bibfield  {journal} {\bibinfo  {journal} {arXiv preprint arXiv:1902.10104}\
  } (\bibinfo {year} {2019})}\BibitemShut {NoStop}%
\bibitem [{\citenamefont {Carleo}\ \emph {et~al.}()\citenamefont {Carleo},
  \citenamefont {Choo}, \citenamefont {Hofmann}, \citenamefont {Smith},
  \citenamefont {Westerhout}, \citenamefont {Alet}, \citenamefont {Davis},
  \citenamefont {Efthymiou}, \citenamefont {Glasser}, \citenamefont {Lin},
  \citenamefont {Mauri}, \citenamefont {Mazzola}, \citenamefont {Mendl},
  \citenamefont {van Nieuwenburg}, \citenamefont {O'Reilly}, \citenamefont
  {Th{\'e}veniaut}, \citenamefont {Torlai},\ and\ \citenamefont
  {Wietek}}]{netket}%
  \BibitemOpen
  \bibfield  {author} {\bibinfo {author} {\bibfnamefont {G.}~\bibnamefont
  {Carleo}}, \bibinfo {author} {\bibfnamefont {K.}~\bibnamefont {Choo}},
  \bibinfo {author} {\bibfnamefont {D.}~\bibnamefont {Hofmann}}, \bibinfo
  {author} {\bibfnamefont {J.~E.~T.}\ \bibnamefont {Smith}}, \bibinfo {author}
  {\bibfnamefont {T.}~\bibnamefont {Westerhout}}, \bibinfo {author}
  {\bibfnamefont {F.}~\bibnamefont {Alet}}, \bibinfo {author} {\bibfnamefont
  {E.~J.}\ \bibnamefont {Davis}}, \bibinfo {author} {\bibfnamefont
  {S.}~\bibnamefont {Efthymiou}}, \bibinfo {author} {\bibfnamefont
  {I.}~\bibnamefont {Glasser}}, \bibinfo {author} {\bibfnamefont {S.-H.}\
  \bibnamefont {Lin}}, \bibinfo {author} {\bibfnamefont {M.}~\bibnamefont
  {Mauri}}, \bibinfo {author} {\bibfnamefont {G.}~\bibnamefont {Mazzola}},
  \bibinfo {author} {\bibfnamefont {C.~B.}\ \bibnamefont {Mendl}}, \bibinfo
  {author} {\bibfnamefont {E.}~\bibnamefont {van Nieuwenburg}}, \bibinfo
  {author} {\bibfnamefont {O.}~\bibnamefont {O'Reilly}}, \bibinfo {author}
  {\bibfnamefont {H.}~\bibnamefont {Th{\'e}veniaut}}, \bibinfo {author}
  {\bibfnamefont {G.}~\bibnamefont {Torlai}}, \ and\ \bibinfo {author}
  {\bibfnamefont {A.}~\bibnamefont {Wietek}},\ }\href@noop {} {\ }\BibitemShut
  {NoStop}%
\bibitem [{\citenamefont {Jones}\ \emph {et~al.}(2001)\citenamefont {Jones},
  \citenamefont {Oliphant}, \citenamefont {Peterson} \emph
  {et~al.}}]{scipy_2001}%
  \BibitemOpen
  \bibfield  {author} {\bibinfo {author} {\bibfnamefont {E.}~\bibnamefont
  {Jones}}, \bibinfo {author} {\bibfnamefont {T.}~\bibnamefont {Oliphant}},
  \bibinfo {author} {\bibfnamefont {P.}~\bibnamefont {Peterson}},  \emph
  {et~al.},\ }\href {http://www.scipy.org/} {\enquote {\bibinfo {title}
  {{SciPy}: Open source scientific tools for {Python}},}\ } (\bibinfo {year}
  {2001})\BibitemShut {NoStop}%
\bibitem [{\citenamefont {Johansson}\ \emph {et~al.}(2013)\citenamefont
  {Johansson}, \citenamefont {Nation},\ and\ \citenamefont {Nori}}]{qutip}%
  \BibitemOpen
  \bibfield  {author} {\bibinfo {author} {\bibfnamefont {J.~R.}\ \bibnamefont
  {Johansson}}, \bibinfo {author} {\bibfnamefont {P.~D.}\ \bibnamefont
  {Nation}}, \ and\ \bibinfo {author} {\bibfnamefont {F.}~\bibnamefont
  {Nori}},\ }\href@noop {} {\bibfield  {journal} {\bibinfo  {journal} {Computer
  Physics Communications}\ }\textbf {\bibinfo {volume} {184}},\ \bibinfo
  {pages} {1234} (\bibinfo {year} {2013})}\BibitemShut {NoStop}%
\end{thebibliography}%
\end{document}